\documentclass[12pt]{article}

\begin{document}

\begin{center}

{\Large {Spinor vertices and bubbles in the "old" conformal bootstrap on AdS and FMH problem: Yukawa model}}

\vspace{1,5cm}

{Boris L. Altshuler}\footnote{E-mail addresses: baltshuler@yandex.ru $\,\,\,  \& \,\,\,$  altshul@lpi.ru}

\vspace{1cm}

{\it Theoretical Physics Department, P.N. Lebedev Physical
Institute, \\  53 Leninsky Prospect, Moscow, 119991, Russia}

\vspace{1,5cm}

\end{center}

{\bf Abstract:} Attempts to resolve the longstanding Fermion mass hierarchy (FMH) problem in frames of the AdS/CFT correspondence demand the knowledge of bulk fermion masses. The approach of the "old" conformal bootstrap in AdS context permitted to calculate bulk masses of scalar fields [1], [2]. In the present paper this approach is extended to physically more interesting spin 1/2 bulk fields. Calculation of spinor-scalar vertices is performed in physical AdS space, and unexpectedly simple expressions for spinor-scalar  bubbles (2-point one-loop self-energy Witten diagrams) are obtained. The "double-trace from UV to IR flow" subtraction of UV divergences used earlier in calculations of the UV-finite bulk tadpoles is applied to bubbles. This permitted to write down in frame of the "old" conformal bootstrap approach the nontrivial spectral equations for the bulk fermion masses; SU(N) Yukawa model of spinor fields interacting with conformal invariant scalar field in case of four boundary dimensions is considered as a specific example.

\vspace{0,5cm}


\newpage

\tableofcontents

\newpage

\section{Introduction. Physical motivation}

\quad  To explain the fermion mass hierarchy (also called flavor hierarchy problem) remains a challenge to theoretical physics (see e.g. \cite{Diaz0}, \cite{Abbas} and references therein). I can't help quoting the very beginning of paper \cite{Abbas}: {\it{"In a recent interview published in CERN COURIER, Steven Weinberg was asked what single open question he would like to see answered in his lifetime, and Weinberg replied that it is only the mystery of the observed pattern of quarks and leptons masses \cite {Weinberg}"}}. 

In frames of the AdS/CFT approach and two-branes Randall-Sundrum model \cite{Randall} (Poincare coordinate $\epsilon < z < L$ where $\epsilon^{-1} \sim M_{Pl} = 10^{19}$ GeV - Planck scale and $L^{-1} \sim M_{EW} = 10^{2}$ GeV - electro-weak scale) spectra of physical particles (glueballs, mesons, light and heavy fermions...) are obtained as eigenvalues of equations for bulk fields, and it is possible in principle to get the looked for fermion masses of intermediate scales with the choice of dynamics in the bulk and of boundary conditions on both branes \cite{Pomarol} - \cite{Archer}. It is evident that there is great arbitrariness in this approach.

Here we mention one of possibilities to overcome this arbitrariness. This is rather natural "twisted" b.c. for spin $1/2$ fields that permit to get observable fermion masses of any scale depending only on the bulk masses of Fermi fields in higher dimensions \cite{Neubert}, \cite{Gher2}. Spectral equation in this case looks as (see formula (29) in \cite{Gher2}):

\begin{equation}
\label{1}
\frac{J_{\alpha - 1}(m_{n}\epsilon)}{Y_{\alpha - 1}(m_{n}\epsilon)} = \frac{J_{\alpha}(m_{n}L)}{Y_{\alpha}(m_{n}L)}, 
\end{equation}
where physical four-momentum $p_{n}^{2} = - m_{n}^{2}$, $J_{\alpha}$, $Y_{\alpha}$ are Bessel functions of the first and second kind of order $\alpha = m/k_{AdS} + 1/2$ ($m$ is bulk mass of fermion field, $k_{AdS}$ is AdS space curvature), values of $\epsilon$ and $L$ are given above. Surely (\ref{1}) has a tower of solutions beginning from the electro-weak scale $L^{-1}$. But it also has a special solution when both arguments of Bessel functions in (\ref{1}) are small:

\begin{equation}
\label{2}
p_{0} = \frac{1}{L}\,\left(\frac{\epsilon}{L}\right)^{\alpha - 1}\,\sqrt{4\alpha(\alpha - 1)}.
\end{equation}
It is seen that for $1 < \alpha < 2$, that is for

\begin{equation}
\label{3}
\frac{1}{2} < \frac{m}{k_{AdS}} < \frac{3}{2}
\end{equation}
$p_{0}$ may have any value in interval from the electro-weak scale $M_{EW}$ to the electron neutrino (or gravitino) mass scale of order $M_{EW}^{2} / M_{Pl}$. Thus well grounded calculations of fermions bulk masses may open the way to solution of the FMH problem. This was the main motivation of the present paper. 

One of the candidates for the theory capable to fix masses of bulk fields may be the "old" conformal bootstrap in the AdS/CFT context \cite{Alt3}, \cite{Alt4}. "Old" conformal bootstrap was proposed about 50 years ago in pioneer papers \cite{old1}, \cite{old2} and developed in \cite{Parisi} - \cite{Dobrev2} (see e.g. \cite{Grensing} and references therein) as a non-Lagrangian tool of self-consistent calculations of conformal dimensions. 

In the AdS/CFT context finding of conformal dimensions is equivalent to finding of bulk masses. In \cite{Alt3} the values of bulk masses of scalar fields that are the roots of corresponding bootstrap spectral equations were found in certain models under the oversimplifying assumption of replacing of two "intermediate" Green functions in self-energy Witten diagrams (bubbles) by their harmonic counterparts. In \cite{Alt4} this assumption was abandoned, and UV divergence of the bubble was subtracted in the "double-trace from UV to IR flow" way used earlier in calculations of the UV-finite tadpoles and bulk vacuum quantum energies of scalar \cite{Mitra}-\cite{Alt2} and spinor \cite{Allais}-\cite{Diaz2} bulk fields. The sensible, that is obeying unitarity bound demand, values of conformal dimensions of scalar fields in $O(N)$ symmetric model were obtained in \cite{Alt4} for $N = 1...4$ in case of $d = 4$ boundary dimensions. 

To extend the approach of \cite{Alt4} to physically more interesting spin $1/2$ bulk fields is the goal of the present paper. 

Dirac field of spin $1/2$ was studied sufficiently well from the AdS/CFT perspective - see earlier papers \cite{Henning} - \cite{Rashkov}, \cite{Allais} where in particular spinor bulk-to-bulk and bulk-to-boundary propagators on AdS were written down, and recent works \cite{Hikida} - \cite{Carmi}. In \cite{Isono}, \cite{Nishida} spinor-spinor-scalar vertices were calculated in formalism of embedding space and also in \cite{Nishida} spectral representation of the bulk spinor Green function is presented. Bulk fermion loop of scalar field was first calculated in \cite{Carmi} also in the formalism of embedding space, whereas one-loop self-energy of Fermi field on AdS was not calculated earlier, as to our knowledge. Here we don't use formalism of embedding space and perform calculations in physical $AdS_{d + 1}$.

In Sec. 2 some well known expressions are presented including four bulk-to-boundary propagators, spectral representation of Green function and split representation of harmonic function for spin 1/2 bulk field. Also subsection 2.3 presents novel bulk and conformal integrals necessary for calculation of spinor-scalar vertices and bubbles; their derivation is given in Appendix.

In Sec 3 calculation of spinor-spinor-scalar 3-point correlators (vertices) is performed. There are two types of such correlators: type $(I)$ where two spinor fields are of one and the same asymptotic at the AdS horizon (symbolically those are correlators $IR-IR-\phi$ or $UV-UV-\phi$, $\phi$ is scalar field), and type $(II)$ - 3-point correlators $IR-UV-\phi$ or $UV-IR-\phi$. The first type of these correlators is well known and was extensively used in calculations of spinor-scalar Witten diagrams \cite{Kawano} - \cite{Carmi}. Whereas vertices of the second type were not written down so far, as to our knowledge.

In Sec. 4 the one-loop quantum contributions (bubbles) to the 2-point boundary-to-boundary conformal correlators of scalar field (loop is formed by two spinor bulk Green functions) and of spinor field (loop is formed by spinor and scalar Green functions) are calculated. The double integral spectral representations are given for both bubbles repeating the approach of \cite{Giombi1} where double integral spectral representations were put down for bubbles formed by the fields of integer spin. The numerators of integrands in these representations are formed like ordinary bubbles but with replacement in their expressions of intermediate Green functions to the corresponding harmonic functions (proportional to the difference of "UV" and "IR" Green functions). Calculations of harmonic bubbles formed by two spinors and by spinor and scalar are the main content of Sec. 4. The reward for these rather lengthy calculations is the simplicity of final formulas for both bubbles expressed through one and the same universal function of conformal dimensions; moreover, for even number $d$ of boundary dimensions this universal function is just a combination of elementary functions. 

In Sec. 5 UV divergence of the bubbles is subtracted with a tool used earlier in calculations of the UV-finite Witten tadpoles \cite{Mitra} - \cite{Diaz2}. The tool is simple, it says that instead of ordinary Witten diagrams the difference of two similar Witten diagrams built of "UV" and "IR" bulk Green functions must be considered. This difference is the deformation in amplitudes under double-trace flow from UV to IR boundary conditions, as it was first proposed in \cite{KlebWitt} (see \cite{Giombi2} and references therein). A general concept is put forward for such a crucial redefinition of quantum amplitudes: it is proposed to construct Witten diagrams using quantum generation functional (\ref{68}) which is the ratio of two standard functionals built with use of "UV" and "IR" boundary conditions; most generally this approach was studied in \cite{Barv2010} where it was shown that in "ratio-functionals" functional integrals over fields' bulk degrees of freedom reduce, and only quantum theory of boundary fields is taken into account. It is shown in the paper that this procedure gives well defined UV-finite expressions for bubble diagrams but whether it will work for triangle and other Witten diagrams is an open question.

In Sec. 6 transparent formulas for spinor-spinor and spinor-scalar bubbles are derived in the $SU(N)$ model with Yukawa interaction of $N$ spinor fields with the conformal invariant scalar field, in case of 4 boundary dimensions.

In Sec. 7 results of Sec. 6 are used to write down the "old" conformal bootstrap spectral equations for bulk fermion mass $m$ in the $SU(N)$ symmetric option when masses of all $N$ spinors are equal. The system of $N$ spectral equations that will permit to study the possibility of spontaneous breakdown of $SU(N)$ symmetry is also presented in Sec. 7.

Conclusion sums up three principle results of the paper and outlines the possible directions of future work.

\section{Preliminaries}

\subsection{Scalar field on AdS.}

\vspace{0,5cm}

We work in Euclidean $AdS_{d+1}$ in Poincare coordinates $Z^{\mu} = \{z_{0}, {\vec z}\}$ ($\mu = 0, 1, ...d$), where AdS curvature $k_{AdS}$ is put equal to one:

\begin{equation}
\label{4}
ds^{2} = \frac{dz_{0}^{2} + d {\vec z}\,^{2}}{z_{0}^{2}},
\end{equation}
and consider bulk scalar and spinor fields. 

Bulk scalar field $\phi (X)$ of mass $M$ is dual to boundary conformal operator $O^{IR}_{\Delta_{\phi}}(\vec x)$ or to its "shadow" operator $O^{UV}_{d - \Delta_{\phi}}(\vec x)$ with scaling dimensions

\begin{equation}
\label{5}
\Delta_{\phi}^{IR} \equiv \Delta_{\phi} = \frac{d}{2} + \nu; \, \, \, \,  \, \nu = \sqrt{\frac{d^{2}}{4} + M^{2}}; \,\,\,\,  \Delta_{\phi}^{UV} = d - \Delta_{\phi}.
\end{equation}

We take normalization of the scalar field's bulk-to-boundary propagator and of the corresponding conformal correlator like in \cite{Giombi1}:

\begin{eqnarray}
\label{6}
K_{\Delta} (Z, \vec x) = \lim_{\stackrel {x_{0} \to 0}{}} \left[\frac{G_{\Delta}^{BB} (Z, X)}{(x_{0})^{\Delta}}\right] = C_{\Delta}\, \cdot \, Q^{\Delta}(Z, \vec x), \qquad \nonumber
\\
\\
 C_{\Delta} = \frac{\Gamma (\Delta)}{2\pi^{d/2}\Gamma \left(1 + \Delta - \frac{d}{2}\right)}, \qquad Q (Z, {\vec x}) = \frac{z_{0}}{z_{0}^{2} + ({\vec z} - {\vec x})^{2}}. \nonumber
\end{eqnarray}
and:

\begin{equation}
\label{7}
<O_{\Delta}({\vec x}) O_{\Delta} ({\vec y})> = \lim_{\stackrel{x_{0} \to 0}{y_{0} \to 0}} \left[\frac{G_{\Delta}^{BB} (X, Y)}{(x_{0}\,y_{0})^{\Delta}}\right]= \frac{C_{\Delta}}{P_{xy}^{\Delta}}, \, \, \,  P_{xy} = |{\vec x} - {\vec y}|^{2}
\end{equation}

Bulk-to-bulk IR ($\Delta = \Delta^{IR} > d/2$, see (\ref{5})) scalar field Green function $G_{\Delta}^{IR} (X, Y)$ possesses Kallen-Lehmann type spectral representation where numerator of the integrand (harmonic function) admits split representation \cite{Penedones} - \cite{Bekaert}, \cite{Giombi1}:

\begin{eqnarray}
\label{8}
G_{\Delta}^{IR} (X, Y) = \int_{-\infty}^{+\infty} \frac{\Omega_{c,0}(X,Y)\,dc}{[c^{2} + (\Delta - \frac{d}{2})^{2}]},  \qquad \qquad \nonumber
\\
\\
\Omega_{c,0}(X,Y) = \frac{c^{2}}{\pi} \, \int \, K_{\frac{d}{2} + ic}(X, {\vec x}_{a})\, K_{\frac{d}{2} - ic}(Y, {\vec x}_{a})\, d^{d}{\vec x}_{a} \, , \nonumber
\end{eqnarray}
where

\begin{equation}
\label{9}
\Omega_{c,0}(X,Y) = \frac{ic}{2\pi}\,{\widetilde G}_{\frac{d}{2} + ic}
\end{equation}
is scalar field harmonic function which is proportional to the difference (marked here with tilde) of IR and UV bulk Green functions:

\begin{equation}
\label{10}
{\widetilde G}_{\Delta}(X, Y) = G_{\Delta}^{IR} - G_{d-\Delta}^{UV} = (d - 2\Delta)\, \int \, K_{\Delta}(X, {\vec x}_{a})\, K_{d-\Delta}(Y, {\vec x}_{a})\, d^{d}{\vec x}_{a}.
\end{equation}

\vspace{0,5cm}

\subsection{Spinor field on AdS.}

\vspace{0,5cm}

Bulk spinor field $\psi(X)$ of mass $m$ on $AdS_{d+1}$ obeys Dirac equation \cite{Henning} - \cite{Carmi}

\begin{equation}
\label{11}
(\gamma^{\mu}\,D_{\mu} - m) \psi(X) = \left(z_{0}\gamma^{0}\frac{\partial}{\partial z_{0}} + z_{0}{\vec \gamma}\frac{\partial}{\partial {\vec z}} - \frac{d}{2}\gamma^{0} - m\right) \,\,\psi (X) = 0,
\end{equation}
$\gamma^{\mu} = \{\gamma^{0}, {\vec \gamma}\}$ are standard anti-commuting gamma matrices in $(d + 1)$ dimensional Euclidean space: $\gamma^{\mu}\gamma^{\nu} + \gamma^{\nu}\gamma^{\mu} = 2\delta^{\mu\nu}$. Bulk spinor field of mass $m$ is dual to boundary conformal operator $O^{IR}_{\Delta_{\psi}} (\vec x)$ or to its "shadow" operator $O^{UV}_{d - \Delta_{\psi}}(\vec x)$ with scaling dimensions

\begin{equation}
\label{12}
\Delta_{\psi}^{IR} = \frac{d}{2} + m + \frac{1}{2} \equiv \Delta_{\psi} + \frac{1}{2}; \qquad \Delta_{\psi}^{UV} = \frac{d}{2} - m + \frac{1}{2} = d - \Delta_{\psi} + \frac{1}{2}.
\end{equation}
Corresponding conformal correlators are:

\begin{eqnarray}
\label{13}
<{\overline O}^{IR}_{\Delta_{\psi}}(\vec x)O^{IR}_{\Delta_{\psi}}(\vec y)> = {\hat C}_{\Delta_{\psi}}\frac{{\vec \gamma}({\vec x} - {\vec y}) \, \Pi_{-}}{P_{xy}^{\Delta_{\psi} + \frac{1}{2}}}, \,\,\,\, {\hat C}_{\Delta_{\psi}} = \frac{\Gamma (\Delta_{\psi} + \frac{1}{2})}{\pi^{d/2}\Gamma \left(\Delta_{\psi} + \frac{1}{2} - \frac{d}{2}\right)}, \nonumber
\\
\\
<{\overline O}^{UV}_{d - \Delta_{\psi}}(\vec x)O^{UV}_{d - \Delta_{\psi}}(\vec y)> = {\hat C}_{d - \Delta_{\psi}}\frac{{\vec \gamma}({\vec x} - {\vec y}) \, \Pi_{+}}{P_{xy}^{d - \Delta_{\psi} + \frac{1}{2}}},  \qquad  \qquad \qquad \nonumber
\end{eqnarray}
where $\Pi_{\pm}$ are projective operators:

\begin{equation}
\label{14}
\Pi_{\pm} = \frac{1}{2}\,(1 \pm \gamma^{0}), \,\,\, \Pi_{+}^{2} = \Pi_{+}, \,\,\, \Pi_{-}^{2} = \Pi_{-}, \,\,\, \Pi_{+}\Pi_{-} = 0, \,\,\, \Pi_{+}{\vec \gamma} = {\vec \gamma}\Pi_{-}.
\end{equation}

Bulk spinor Green functions $S^{IR,UV}(X, Y)$ that possess $IR$ or $UV$ asymptotic at the horizon, $x_{0}, y_{0} \to 0$, and that are zero at the AdS infinity, $x_{0}, y_{0} \to \infty$, are well known, see e.g. in \cite{Allais}, \cite{Kawano}, \cite{Rashkov}. Their properly normalized limits at the horizon give four spinor bulk-to-boundary propagators $\Sigma^{IR(UV)}$ and ${\overline\Sigma}^{IR(UV)}$, which permit to express bulk solutions of Dirac equation (\ref{11}) $\psi(Z)$, ${\overline \psi}(Z)$ through the boundary fields \cite{Henning} - \cite{Rashkov}. Those bulk-to-boundary propagators are the main tool in calculations of this paper:

\begin{equation}
\label{15}
\Sigma^{IR}_{\Delta_{\psi}}(Z,{\vec y}) = \lim_{\stackrel {y_{0} \to 0}{}}\left[ \frac{S^{IR}_{\Delta_{\psi}} (Z, Y) }{y_{0}^{\Delta_{\psi} + \frac{1}{2}}}\right] = {\hat C}_{\Delta_{\psi}}\,Q^{\Delta_{\psi} + \frac{1}{2}}(Z, {\vec y})\, \frac{[z_{0} - {\vec \gamma}({\vec z} - {\vec y})]}{\sqrt z_{0}}\,\Pi_{-};
\end{equation}

\begin{equation}
\label{16}
{\overline \Sigma}^{IR}_{\Delta_{\psi}}(Z,{\vec x}) = \lim_{\stackrel {x_{0} \to 0}{}}\left[ \frac{S^{IR}_{\Delta_{\psi}} (X, Z) }{x_{0}^{\Delta_{\psi} + \frac{1}{2}}}\right] = {\hat C}_{\Delta_{\psi}}\,Q^{\Delta_{\psi} + \frac{1}{2}}(Z, {\vec x})\,\Pi_{+} \, \frac{[z_{0} + {\vec \gamma}({\vec z} - {\vec x})]}{\sqrt z_{0}};
\end{equation}
and in a similar way for $\Sigma^{UV}$ and ${\overline \Sigma}^{UV}$:

\begin{equation}
\label{17}
\Sigma^{UV}_{d - \Delta_{\psi}}(Z,{\vec y}) = {\hat C}_{d - \Delta_{\psi}}\,Q^{d - \Delta_{\psi} + \frac{1}{2}}(Z, {\vec y})\, \frac{[z_{0} + {\vec \gamma}({\vec z} - {\vec y})]}{\sqrt z_{0}}\,\Pi_{+};
\end{equation}

\begin{equation}
\label{18}
{\overline \Sigma}^{UV}_{d - \Delta_{\psi}}(Z,{\vec x}) = {\hat C}_{d - \Delta_{\psi}}\,Q^{d - \Delta_{\psi} + \frac{1}{2}}(Z, {\vec x})\,\Pi_{-} \, \frac{[z_{0} - {\vec \gamma}({\vec z} - {\vec x})]}{\sqrt z_{0}},
\end{equation}
where ${\hat C}_{\Delta_{\psi}}$ and $Q(Z, {\vec x})$ are defined correspondingly in (\ref{13}) and (\ref{6}).

Conformal spinor correlators (\ref{13}) are obtained from (\ref{15})-(\ref{18}) when bulk coordinate is sent to horizon in these expressions.

Analogous to scalar case (\ref{8}) spectral representation for $S^{IR}_{\Delta_{\psi}} (X, Y)$ ($\Delta_{\psi} = d/2 + m > d/2$, see (\ref{12})) was given in \cite{Nishida}:

\begin{eqnarray}
\label{19}
S^{IR}_{\Delta_{\psi}} (X, Y) = \int_{-\infty}^{+\infty} \frac{\Omega_{c,1/2}(X,Y)\,dc}{[c + i\,(\Delta_{\psi} - \frac{d}{2})]}, \qquad \nonumber
\\
\\
\Omega_{c,1/2}(X,Y) = \frac{i}{2\pi}\, {\widetilde S}_{\Delta_{c}}(X, Y), \qquad \Delta_{c} = \frac{d}{2} + i\,c,  \nonumber  
\end{eqnarray}
where spinor harmonic function $\Omega_{c,1/2}(X,Y)$ is proportional to the difference (again marked by tilde) of $IR$ and $UV$ Green functions, that possesses nice split representations through bulk-to-boundary propagators (\ref{15}) - (\ref{18}) integrated over common boundary point \cite{Nishida} (cf. (\ref{9}), (\ref{10}) for the scalar field case):

\begin{eqnarray}
\label{20}
{\widetilde S}_{\Delta_{\psi}}(X, Y) = S^{IR}_{\Delta_{\psi}} (X, Y) - S^{UV}_{d-\Delta_{\psi}} (X, Y) = \qquad  \qquad \nonumber
\\
\\
= \int\,\Sigma^{IR}_{\Delta_{\psi}}(X, {\vec x}_{a})\,{\overline \Sigma}^{UV}_{d - \Delta_{\psi}}(Y,{\vec x}_{a})\,d{\vec x}_{a} = \int\,\Sigma^{UV}_{d - \Delta_{\psi}}(X, {\vec x}_{a})\,{\overline \Sigma}^{IR}_{\Delta_{\psi}}(Y,{\vec x}_{a})\,d{\vec x}_{a}.  \nonumber
\end{eqnarray}

\vspace{0,5cm}

\subsection{Some integrals.}

\vspace{0,5cm}

We shall need two following bulk integrals for two values of parameter $\alpha$ ($\alpha = 0$ and $\alpha = 1$):

\begin{equation}
\label{21}
{\rm D}^{(\alpha)}_{\gamma_{1},\gamma_{2},\gamma_{3}}({\vec x_{1}},{\vec x_{2}},{\vec x_{3}}) = \int\,Q^{\gamma_{1}}(Z, {\vec x}_{1})\,Q^{\gamma_{2}}(Z, {\vec x}_{2})\,Q^{\gamma_{3}}(Z, {\vec x}_{3})\,\frac{dZ}{z_{0}^{\alpha}},
\end{equation}
$Q(Z, {\vec x})$ see in (\ref{6}). ${\rm D}^{(0)}$ is well known, it gives the 3-pt scalar fields vertex \cite{Freedman}, \cite{Penedones}, \cite{Paulos}:

\begin{eqnarray}
\label{22}
{\rm D}^{(0)}_{\gamma_{1},\gamma_{2},\gamma_{3}}({\vec x}_{1}, {\vec x}_{2}, {\vec x}_{3}) = \frac{{\rm d}^{(0)}(\gamma_{1},\gamma_{2},\gamma_{3})}{P_{12}^{\delta_{12}}\,P_{13}^{\delta_{13}}\, P_{23}^{\delta_{23}}}, \qquad  \qquad  \nonumber
\\
\\
{\rm d}^{(0)}(\gamma_{1},\gamma_{2},\gamma_{3}) = \frac{\pi^{d/2}}{2} \, \frac{\Gamma \left(\frac{\Sigma \gamma_{i} - d}{2}\right)\, \Gamma(\delta_{12})\, \Gamma(\delta_{13}) \, \Gamma (\delta_{23})}{\Gamma (\gamma_{1})\, \Gamma (\gamma_{2}) \, \Gamma (\gamma_{3})} \nonumber
\end{eqnarray}
where

\begin{equation}
\label{23}
\delta_{12} = \frac{\gamma_{1} + \gamma_{2} - \gamma_{3}}{2}; \,\,\, \delta_{13} = \frac{\gamma_{1} + \gamma_{3} - \gamma_{2}}{2}; \,\,\, \delta_{23} = \frac{\gamma_{2} + \gamma_{3} - \gamma_{1}}{2}.
\end{equation}

Whereas ${\rm D}^{(1)}$ is derived in Appendix, it will be used in calculation of spinor-spinor-scalar vertex of type $\rm II$ in subsection 3.2:

\begin{eqnarray}
\label{24}
{\rm D}^{(1)}_{\gamma_{1},\gamma_{2},\gamma_{3}}({\vec x}_{1}, {\vec x}_{2}, {\vec x}_{3}) = \frac{{\rm d}^{(1)}(\gamma_{1},\gamma_{2},\gamma_{3})}{P_{12}^{{\hat \delta}_{12}}\,P_{13}^{{\hat \delta}_{13}}\, P_{23}^{{\hat \delta}_{23}}} \, \left[\frac{{\hat \delta}_{12}\,{\hat \delta}_{13}}{P_{12} \, P_{13}} + \frac{{\hat \delta}_{12}\,{\hat \delta}_{23}}{P_{12} \, P_{23}} + \frac{{\hat \delta}_{13}\,{\hat \delta}_{23}}{P_{13} \, P_{23}}\right] \nonumber
\\
\\
{\rm d}^{(1)}(\gamma_{1},\gamma_{2},\gamma_{3}) = \frac{\pi^{d/2}}{2} \, \frac{\Gamma \left(\frac{\Sigma \gamma_{i} - d - 1}{2}\right)\, \Gamma({\hat \delta}_{12})\, \Gamma({\hat \delta}_{13}) \, \Gamma ({\hat \delta}_{23})}{\Gamma (\gamma_{1})\, \Gamma (\gamma_{2}) \, \Gamma (\gamma_{3})} \qquad  \qquad  \nonumber
\\  \nonumber
\\ \nonumber
{\hat \delta}_{ij} = \delta_{ij} - \frac{1}{2}. \qquad  \qquad  \qquad  \qquad   \qquad  \qquad   \nonumber
\end{eqnarray}

There are relations between coefficients ${\rm d}^{(0)}$ and ${\rm d}^{(1)}$ that permit significantly simplify the calculations:

\begin{equation}
\label{25} 
{\rm d}^{(0)}(\gamma_{1} - 1,\gamma_{2},\gamma_{3}) = {\rm d}^{(1)}(\gamma_{1},\gamma_{2},\gamma_{3}) \, (\gamma_{1} - 1) \, {\hat \delta}_{23},
\end{equation}
and the same for two other arguments.

Along with the well known conformal integrals \cite{Symanzik2}, \cite{Parisi}, \cite{Fradkin}:

\begin{equation}
\label{26}
R^{(0)}_{\beta_{1}\beta_{2}\beta_{3}}({\vec x}_{1}, {\vec x}_{2}, {\vec x}_{3}) = \int \frac{d^{d}{\vec y}}{P_{1y}^{\beta_{1}} \,P_{2y}^{\beta_{2}}\, P_{3y}^{\beta_{3}}} \stackrel {\Sigma \beta_{i} = d} {=} \frac{A(\beta_{1}, \beta_{2}, \beta_{3})}{P_{12}^{\frac{d}{2} - \beta_{3}}\,P_{13}^{\frac{d}{2} - \beta_{2}}\, P_{23}^{\frac{d}{2} - \beta_{1}}},
\end{equation}
and

\begin{equation}
\label{27}
\int \frac{d^{d}{\vec y}}{P_{1y}^{\beta_{1}} \,P_{2y}^{\beta_{2}}} = \frac{A(\beta_{1}, \beta_{2}, d - \beta_{1} - \beta_{2})} {P_{12}^{\beta_{1} + \beta_{2} - \frac{d}{2}}},
\end{equation}
where

\begin{equation}
\label{28}
A(\beta_{1}, \beta_{2}, \beta_{3}) = \frac{\pi^{d/2}\, \Gamma (\frac{d}{2} - \beta_{1}) \, \Gamma (\frac{d}{2} - \beta_{2}) \, \Gamma (\frac{d}{2} - \beta_{3})}{\Gamma (\beta_{1}) \, \Gamma(\beta_{2}) \, \Gamma (\beta_{3})},
\end{equation}
the knowledge of two following integrals is necessary for calculation of spinor one-loop self-energy in subsection 4.2:

\begin{eqnarray}
\label{29}
R^{(1)}_{\beta_{1}\beta_{2}\beta_{3}}({\vec x}_{1}, {\vec x}_{2}, {\vec x}_{3}) = \int \frac{d^{d}{\vec y}}{P_{1y}^{\beta_{1}} \,P_{2y}^{\beta_{2}}\, P_{3y}^{\beta_{3}}} \stackrel {\Sigma \beta_{i} = d + 1} {=} \frac{A(\beta_{1}, \beta_{2}, \beta_{3})}{P_{12}^{\frac{d}{2} - \beta_{3}}\,P_{13}^{\frac{d}{2} - \beta_{2}}\, P_{23}^{\frac{d}{2} - \beta_{1}}} \, \cdot   \quad \nonumber
\\
\\
\cdot \, \left[\frac{\left(\frac{d}{2} - \beta_{2}\right)\,\left(\frac{d}{2} - \beta_{3}\right)}{P_{12}\,P_{13}}\,\,+ \,\, \frac{\left(\frac{d}{2} - \beta_{1}\right)\,\left(\frac{d}{2} - \beta_{2}\right)}{P_{13}\,P_{23}} \,\, + \,\, \frac{\left(\frac{d}{2} - \beta_{1}\right)\,\left(\frac{d}{2} - \beta_{3}\right)}{P_{12}\,P_{23}} \right], \nonumber
\end{eqnarray}
and

\begin{eqnarray}
\label{30}
\int d{\vec x}_{a}\,\frac{{\vec \gamma}({\vec x}_{2} - {\vec x}_{a})}{P_{1a}^{\beta_{1}}\,P_{2a}^{\beta_{2}}} = \frac{{\vec \gamma}({\vec x}_{1} - {\vec x}_{2})}{P_{12}^{\beta_{1} + \beta_{2} - \frac{d}{2}}}\, \cdot \, \frac{1}{2}\,[A(\beta_{1} - 1, \beta_{2}, d - \beta_{1} - \beta_{2} + 1) -  \nonumber
\\
\\
- A(\beta_{1}, \beta_{2} - 1, d - \beta_{1} - \beta_{2} +1) - A(\beta_{1}, \beta_{2}, d - \beta_{1} - \beta_{2})]. \qquad  \nonumber
\end{eqnarray}

Derivation of (\ref{29}) is given in Appendix. Whereas expression (\ref{30}) it's easy to prove by multiplying it by ${\vec \gamma}({\vec x}_{1} - {\vec x}_{2})$ with account of (\ref{27}) and identity:

\begin{eqnarray}
\label{31}
{\vec \gamma}({\vec x}_{1} - {\vec x}_{2}) \,\cdot \, {\vec \gamma}({\vec x}_{2} - {\vec x}_{a}) = \frac{1}{2}\,[P_{1a} - P_{2a} - P_{12}] + ({\vec x_{1}} - {\vec x_{2}})^{\alpha} \, ({\vec x_{2}} - {\vec x_{a}})^{\beta} \, S^{\alpha\beta}, \nonumber
\\
\\
S^{\alpha\beta} = \frac{\gamma^{\alpha}\gamma^{\beta} - \gamma^{\beta} \gamma^{\alpha}}{2}. \qquad \, \qquad \qquad  \qquad  \qquad \qquad \nonumber
\end{eqnarray}
Because of the $\alpha \leftrightarrow \beta$ asymmetry of the $S^{\alpha\beta}$ term in (\ref{31}) its contribution to the RHS of (\ref{30}) is zero.

\section{Spinor-spinor-scalar vertices}

\subsection{Spinor-spinor-scalar vertex: type ${\rm I}$.}

\vspace{0,5cm}

3-point correlators of scalar field $\phi$ of conformal dimension $\Delta_{\phi}$ and of two spinor fields $\psi_{1}$, $\psi_{2}$, generally speaking of different conformal dimensions $\Delta^{IR(UV)}_{\psi_{1}}$, $\Delta^{IR(UV)}_{\psi_{2}}$ (see (\ref{12})) are generated by the bulk Yukawa interaction ${\cal L}_{int} = g \cdot {\overline \psi_{1}}(Z)\,\psi_{2}(Z)\,\phi(Z)$.

These 3-pt correlators are of two essentially different types, as regards to spinor fields: (I) of coinciding - ${\overline {IR}}-IR$, or ${\overline {UV}}-UV$, and (II) of the opposite - ${\overline {IR}}-UV$ or ${\overline {UV}}-IR$ asymptotics of spinor tails. 

Vertex ${\overline {IR}}-IR$ of the first type

\begin{eqnarray}
\label{32}
{\cal M}^{{\rm 3pt\,(I)}\,{\overline {IR}}-IR}_{\Delta_{\psi_{1}},\Delta_{\psi_{2}},\Delta_{\phi}}({\vec x_{1}},{\vec x_{2}},{\vec x_{3}}) = <{\overline O}^{IR}_{\Delta_{\psi_{1}}}({\vec x}_{1})\,O^{IR}_{\Delta_{\psi_{2}}}({\vec x}_{2})\,O_{\Delta_{\phi}}({\vec x}_{3}) > =  \nonumber
\\
\\
= \int_{AdS} dZ\,{\overline \Sigma}^{IR}_{\Delta_{\psi_{1}}}(Z,{\vec x}_{1})\, \Sigma^{IR}_{\Delta_{\psi_{2}}}(Z,{\vec x}_{2}) \, K_{\Delta_{\phi}}(Z, {\vec x}_{3})  \qquad  \nonumber
\end{eqnarray}
is quite simple because its dependence on the bulk coordinates $(z_{0}, {\vec z})$ drops out from the spinor numerator in (\ref{32}). Really, according to (\ref{15}), (\ref{16}):

\begin{eqnarray}
\label{33}
{\overline \Sigma}^{IR}_{\Delta_{\psi_{1}}}(Z,{\vec x}_{1})\, \Sigma^{IR}_{\Delta_{\psi_{2}}}(Z,{\vec x}_{2}) \sim  \qquad \qquad  \qquad  \qquad \nonumber
\\
\\
\sim \frac{\Pi_{+} \,[z_{0} + {\vec \gamma}({\vec z} - {\vec x}_{1})]}{\sqrt z_{0}} \, \cdot \, \frac{[z_{0} - {\vec \gamma}({\vec z} - {\vec x}_{2})]\,\Pi_{-}}{\sqrt z_{0}} = {\vec \gamma}({\vec x}_{2} - {\vec x}_{1})\,\Pi_{-}   \nonumber
\end{eqnarray}
(we remind that $\Pi_{+} \,\Pi_{-} = 0$, $\Pi_{+} \,{\vec \gamma}\,\Pi_{-} = {\vec \gamma}\,\Pi_{-}$).
In the same way in the ${\overline {UV}}-UV$ case:

\begin{equation}
\label{34}
{\overline \Sigma}^{UV}_{d - \Delta_{\psi_{1}}}(Z,{\vec x}_{1})\, \Sigma^{UV}_{d - \Delta_{\psi_{2}}}(Z,{\vec x}_{2}) \sim {\vec \gamma}({\vec x}_{1} - {\vec x}_{2})\,\Pi_{+}.
\end{equation}

After substitution in (\ref{32}) of three bulk-to-boundary propagators from (\ref{15}), (\ref{16}), (\ref{6}) with account of (\ref{33}) it is obtained:

\vspace{1,5cm}

\begin{eqnarray}
\label{35}
{\cal M}^{{\rm 3 pt \, (I)}\,{\overline {IR}}-IR}_{\Delta_{\psi_{1}},\Delta_{\psi_{2}},\Delta_{\phi}}({\vec x_{1}},{\vec x_{2}},{\vec x_{3}}) = \qquad  \qquad \qquad  \qquad \nonumber
\\
\\
= \left(\prod\limits_{i=1}^{2}{\hat C}_{\Delta_{\psi_{i}}}\right)\,C_{\Delta_{\phi}}\, {\rm D}^{(0)}_{\Delta_{\psi_{1}} +\frac{1}{2},\Delta_{\psi_{2}} + \frac{1}{2},\Delta_{\phi}}({\vec x_{1}},{\vec x_{2}},{\vec x_{3}})\,{\vec \gamma}({\vec x}_{2} - {\vec x}_{1})\,\Pi_{-} =  \nonumber
\\   \nonumber
\\   \nonumber
= B^{\rm (I)}(\Delta_{\psi_{1}}, \Delta_{\psi_{2}},\Delta_{\phi}; 1/2) \cdot \frac{{\vec \gamma}({\vec x}_{2} - {\vec x}_{1})\,\Pi_{-}}{P_{12}^{\delta_{12}^{\rm (I)}}\,P_{13}^{\delta_{13}^{\rm (I)}}\,P_{23}^{\delta_{23}^{\rm (I)}}\,}, \qquad  \qquad \qquad  
\end{eqnarray}
where (see (\ref{22}) for ${\rm D}^{(0)}$):

\begin{eqnarray}
\label{36}
B^{\rm (I)}(\Delta_{\psi_{1}}, \Delta_{\psi_{2}},\Delta_{\phi}; 1/2) = \frac{{\hat C}_{\Delta_{\psi_{1}}}}{\Gamma(\Delta_{\psi_{1}} + \frac{1}{2})}\,\frac{{\hat C}_{\Delta_{\psi_{2}}}}{\Gamma(\Delta_{\psi_{2}} + \frac{1}{2})}\,\frac{C_{\Delta_{\phi}}}{\Gamma(\Delta_{\phi})} \, \cdot \qquad  \qquad  \nonumber
\\
\\ 
\cdot \, \frac{\pi^{d/2}}{2}\,\Gamma\left(\frac{\Delta_{\psi_{1}} + \Delta_{\psi_{2}} + \Delta_{\phi} + 1 - d}{2}\right)\, \Gamma(\delta_{12}^{\rm (I)})\, \Gamma(\delta_{13}^{\rm (I)})\, \Gamma(\delta_{23}^{\rm (I)}), \qquad  \qquad  \nonumber
\\    \nonumber
\\   \nonumber
\delta_{12}^{\rm (I)} = \frac{\Delta_{\psi_{1}}+ \Delta_{\psi_{2}}+ 1 - \Delta_{\phi}}{2};\,\,
\delta_{13}^{\rm (I)} = \frac{\Delta_{\psi_{1}} - \Delta_{\psi_{2}} + \Delta_{\phi}}{2}; \,\,
\delta_{23}^{\rm (I)} = \frac{\Delta_{\psi_{2}} - \Delta_{\psi_{1}} + \Delta_{\phi}}{2}.  \nonumber
\end{eqnarray}

Formula similar to (\ref{36}) is obtained for ${\cal M}^{{\rm 3pt (I)}\,{\overline {UV}}-UV}_{d - \Delta_{\psi_{1}},d - \Delta_{\psi_{2}},\Delta_{\phi}}({\vec x_{1}},{\vec x_{2}},{\vec x_{3}})$ with account of (\ref{34}) and with replacements  $\Delta_{\psi_{1,2}} \to d - \Delta_{\psi_{1,2}}$ in (\ref{35}), (\ref{36}), and ${\vec \gamma}({\vec x}_{2} - {\vec x}_{1})\,\Pi_{-} \to {\vec \gamma}({\vec x}_{1} - {\vec x}_{2})\,\Pi_{+}$ in spinor numerator in the RHS of (\ref{35}).

Simple result (\ref{35}) for ${\overline {IR}}-IR$ spinor-spinor-scalar vertex (of type ${\rm I}$) was obtained in \cite{Kawano} in physical $AdS_{d + 1}$ space and in \cite{Hikida} - \cite{Nishida} in formalism of embedding space.

\vspace{0,5cm}

\subsection{Spinor-spinor-scalar vertex: type ${\rm II}$.}

\vspace{0,5cm}

Let us consider ${\overline {IR}}-UV$ vertex

\begin{eqnarray}
\label{37}
{\cal M}^{{\rm 3pt \, (II)}\,{\overline {IR}}-UV}_{\Delta_{\psi_{1}},d - \Delta_{\psi_{2}},\Delta_{\phi}}({\vec x_{1}},{\vec x_{2}},{\vec x_{3}}) = <{\overline O}^{IR}_{\Delta_{\psi_{1}}}({\vec x}_{1})\,O^{UV}_{d - \Delta_{\psi_{2}}}({\vec x}_{2})\,O_{\Delta_{\phi}}({\vec x}_{3}) > =  \nonumber
\\
\\
= \int_{AdS} dZ\,{\overline \Sigma}^{IR}_{\Delta_{\psi_{1}}}(Z,{\vec x}_{1})\, \Sigma^{UV}_{d - \Delta_{\psi_{2}}}(Z,{\vec x}_{2}) \, K_{\Delta_{\phi}}(Z, {\vec x}_{3}).\qquad  \nonumber
\end{eqnarray}

In this case spinor numerator in (\ref{37}) differs from the one in (\ref{33}) and essentially depends on $z_{0}, {\vec z}$:

\begin{eqnarray}
\label{38}
{\overline \Sigma}^{IR}_{\Delta_{\psi_{1}}}(Z,{\vec x}_{1})\, \Sigma^{UV}_{d - \Delta_{\psi_{2}}}(Z,{\vec x}_{2}) \sim  \frac{\Pi_{+} \,[z_{0} + {\vec \gamma}({\vec z} - {\vec x}_{1})]}{\sqrt z_{0}} \, \cdot \, \frac{[z_{0} + {\vec \gamma}({\vec z} - {\vec x}_{2})]\Pi_{+}}{\sqrt z_{0}} =   \nonumber
\\  \nonumber
\\
\\
= \left\{\frac{1}{2}\,\left[- \frac{P_{12}}{z_{0}} + Q^{-1}(Z,{\vec x}_{1}) + Q^{-1}(Z, {\vec x}_{2}) \right] + \frac{({\vec z} - {\vec x_{1}})^{\alpha} \, ({\vec z} - {\vec x_{2}})^{\beta} \, S^{\alpha\beta}}{z_{0}}\right\} \, \Pi_{+}, \nonumber
\end{eqnarray}
where $P_{12} \equiv P_{x_{1}x_{2}}$, $Q(Z,{\vec x})$ see in (\ref{6}), and $S^{\alpha \beta}$ in (\ref{31}).

The same expression, with the only change of $\Pi_{+} \to \Pi_{-}$ in the RHS of (\ref{38}), is valid for spinor numerator of ${\overline \Sigma}^{UV}_{d - \Delta_{\psi_{1}}}(Z,{\vec x}_{1})\, \Sigma^{IR}_{\Delta_{\psi_{2}}}(Z,{\vec x}_{2})$.

Substitution in (\ref{37}) of the bulk-to-boundary propagators from (\ref{16}), (\ref{17}), (\ref{6}) with account of (\ref{38}) gives for spinor-spinor-scalar vertex of type II:

\begin{eqnarray}
\label{39}
{\cal M}^{{\rm 3pt \, (II)}\,{\overline {IR}}-UV}_{\Delta_{\psi_{1}},d - \Delta_{\psi_{2}},\Delta_{\phi}}({\vec x_{1}},{\vec x_{2}},{\vec x_{3}}) = \qquad  \qquad  \qquad  \qquad  \qquad  \qquad \nonumber
\\   \nonumber
\\   \nonumber
= \frac{1}{2} {\hat C}_{\Delta_{\psi_{1}}}{\hat C}_{d - \Delta_{\psi_{2}}}C_{\Delta_{\phi}}\, \left[- P_{12}\, {\rm D}^{(1)}_{\Delta_{\psi_{1}} +\frac{1}{2},\,d - \Delta_{\psi_{2}} + \frac{1}{2},\,\Delta_{\phi}}({\vec x_{1}},{\vec x_{2}},{\vec x_{3}}) +\right.  \qquad  \qquad   \nonumber
\\   \nonumber
\\  \nonumber
+ {\rm D}^{(0)}_{\Delta_{\psi_{1}} - \frac{1}{2}, d - \Delta_{\psi_{2}} + \frac{1}{2}, \Delta_{\phi}}({\vec x_{1}},{\vec x_{2}},{\vec x_{3}}) + {\rm D}^{(0)}_{\Delta_{\psi_{1}} + \frac{1}{2}, d - \Delta_{\psi_{2}} - \frac{1}{2}, \Delta_{\phi}}({\vec x_{1}},{\vec x_{2}},{\vec x_{3}}) + \qquad 
\\
\\  \nonumber
\left. +  \frac{\pi^{d/2}}{2}\,\frac{\Gamma\left(\frac{\Delta_{\psi_{1}} - \Delta_{\psi_{2}} + \Delta_{\phi}}{2}\right)\, \Gamma(\delta_{12}^{\rm (II)})\, \Gamma(\delta_{13}^{\rm (II)})\, \Gamma(\delta_{23}^{\rm (II)})}{\Gamma(\Delta_{\psi_{1}} + \frac{1}{2}) \, \Gamma(d - \Delta_{\psi_{2}} + \frac{1}{2}) \, \Gamma(\Delta_{\phi})} \, \cdot \, {\vec x_{13}}^{\alpha} \, {\vec x_{23}}^{\beta} \, S^{\alpha\beta} \right] \, \Pi_{+}, \qquad  \nonumber
\end{eqnarray}
${\vec x_{13}}^{\alpha} = ({\vec x_{1}} - {\vec x_{3}})^{\alpha}$, ${\vec x_{23}}^{\beta} = ({\vec x_{2}} - {\vec x_{3}})^{\beta}$; $\delta_{ij}^{(\rm {II})}$ are defined in (\ref{42}) below; derivation of the $S^{\alpha\beta}$ term in the last line of (\ref{39}) is similar to derivation of (\ref{24}), (\ref{29}). 

Then with account of expressions (\ref{22}), (\ref{24}) for ${\rm D}^{(0)}$, ${\rm D}^{(1)}$ and their relations (\ref{25}) final rather simple formula for ${\cal M}^{{\rm 3pt \, (II)}\,{\overline {IR}}-UV}$ is obtained:

\begin{eqnarray}
\label{40}
{\cal M}^{{\rm 3pt \, (II)}\,{\overline {IR}}-UV}_{\Delta_{\psi_{1}},d - \Delta_{\psi_{2}},\Delta_{\phi}}({\vec x_{1}},{\vec x_{2}},{\vec x_{3}}) = B^{\rm (II)}(\Delta_{\psi_{1}},d - \Delta_{\psi_{2}},\Delta_{\phi}; 1/2) \, \cdot \nonumber  
\\
\\
\cdot \, \frac{ - P_{12} + P_{13} + P_{23} + ({\vec x_{1}} - {\vec x_{3}})^{\alpha} \, ({\vec x_{2}} - {\vec x_{3}})^{\beta} \, S^{\alpha\beta}}{P_{12}^{\delta_{12}^{\rm (II)}}\,P_{13}^{\delta_{13}^{\rm (II)}}\,P_{23}^{\delta_{23}^{\rm (II)}}}  \, \Pi_{+}, \qquad \nonumber
\end{eqnarray}
where

\begin{eqnarray}
\label{41}
B^{\rm (II)}(\Delta_{\psi_{1}},d - \Delta_{\psi_{2}},\Delta_{\phi}; 1/2) = \frac{{\hat C}_{\Delta_{\psi_{1}}}}{\Gamma(\Delta_{\psi_{1}} + \frac{1}{2})}\,\frac{{\hat C}_{d - \Delta_{\psi_{2}}}}{\Gamma(d - \Delta_{\psi_{2}} + \frac{1}{2})}\,\frac{C_{\Delta_{\phi}}}{\Gamma(\Delta_{\phi})}\, \cdot \qquad \nonumber
\\
\\
\cdot \, \frac{\pi^{d/2}}{4}\,\Gamma\left(\frac{\Delta_{\psi_{1}} - \Delta_{\psi_{2}} + \Delta_{\phi}}{2}\right)\, \Gamma(\delta_{12}^{\rm (II)})\, \Gamma(\delta_{13}^{\rm (II)})\, \Gamma(\delta_{23}^{\rm (II)}), \qquad \qquad  \qquad  \nonumber
\end{eqnarray}

and

\begin{eqnarray}
\label{42}
\delta_{12}^{\rm (II)} = \frac{\Delta_{\psi_{1}} + d - \Delta_{\psi_{2}} - \Delta_{\phi}}{2};\,\,
\delta_{13}^{\rm (II)} = \frac{\Delta_{\psi_{1}} + \Delta_{\phi} - (d - \Delta_{\psi_{2}}) + 1}{2}; \,\, \nonumber
\\
\\
\delta_{23}^{\rm (II)} = \frac{(d- \Delta_{\psi_{2}}) + \Delta_{\phi} - \Delta_{\psi_{1}} + 1}{2}. \qquad \qquad  \qquad \nonumber
\end{eqnarray}

Expression for ${\cal M}^{{3pt \rm (II)}\,{\overline {UV}}-IR}_{d - \Delta_{\psi_{1}},\Delta_{\psi_{2}},\Delta_{\phi}}({\vec x_{1}},{\vec x_{2}},{\vec x_{3}})$ is obtained from (\ref{40})-(\ref{42}) with the simple replacements $\Delta_{\psi_{1}} \to d - \Delta_{\psi_{1}}$, $d - \Delta_{\psi_{2}} \to \Delta_{\psi_{2}}$ together with $\Pi_{+} \to \Pi_{-}$ in the RHS of (\ref{40}).

\section{Spinor "harmonic bubbles"}

\subsection{Scalar bubble formed by two spinors.}

\vspace{0,5cm}

Fermionic bubble diagram of scalar field on AdS was first calculated in \cite{Carmi} in formalism of embedding space. This one-loop contribution to the two-point correlator of scalar field $\phi(Z)$ is generated by its bulk coupling $g\,\phi(Z){\overline {\psi}}(Z)\chi(Z)$ with two spinor fields; it is formed by the bulk-to-boundary propagators of scalar field $K_{\phi}$ (\ref{6}) and bulk Green functions of spinor fields $S_{\psi}$, $S_{\chi}$ (\ref{19}):

\begin{equation}
\label{43}
{\cal M}^{{\rm 2pt}(0|\frac{1}{2}\,\frac{1}{2})}_{\Delta_{\phi}|\Delta_{\psi}\Delta_{\chi}}({\vec x}_{1}, {\vec x}_{2}) = g^{2} \int\int K_{\Delta_{\phi}}(X ; {\vec x}_{1})\,{\rm Tr}\,[S_{\Delta_{\psi}}(X, Y) \, S_{\Delta_{\chi}} (Y, X)]\,K_{\Delta_{\phi}}(Y ; {\vec x}_{2})
\end{equation}
(trace ${\rm Tr}$ is over spinor indices, bulk integrals over $X$, $Y$ are supposed).

Following approach of \cite{Giombi1} where double integral spectral representations of bubbles of fields of any integer spin were considered and referring to the spectral representation (\ref{19}) of spinor Green function the double integral spectral representation of bubble (\ref{43}) may be put down:

\begin{equation}
\label{44}
{\cal M}^{{\rm 2pt}(0|\frac{1}{2}\,\frac{1}{2})}_{\Delta_{\phi}|\Delta_{\psi}\Delta_{\chi}}({\vec x}_{1}, {\vec x}_{2}) = - \frac{1}{4\pi^{2}}\,\int \int \, \frac{dc\, d{\overline c} \, {\cal H}^{{\rm 2pt}(0|\frac{1}{2}\,\frac{1}{2})}_{\Delta_{\phi}|\frac{d}{2} + ic,\frac{d}{2} + i {\overline c}}\,({\vec x}_{1}, {\vec x}_{2})}{[c + i\,(\Delta_{\psi} - \frac{d}{2})] \, [{\overline c} + i\,(\Delta_{\chi} - \frac{d}{2})]},
\end{equation}
where, taking into account the proportionality of spinor harmonic function $\Omega_{c, \frac{1}{2}}$ entering spectral representation (\ref{19}) to the difference of Green functions ${\widetilde S}$ (\ref{20}), we introduced in numenator of integrand in (\ref{44}) the "harmonic bubble" ${\cal H}$ that is built by the replacement in (\ref{43}) of two bulk spinor Green functions with the corresponding differences ${\widetilde S}$ (\ref{20}):

\begin{equation}
\label{45}
{\cal H}^{{\rm 2pt}(0|\frac{1}{2}\,\frac{1}{2})}_{\Delta_{\phi}|\Delta_{\psi}\Delta_{\chi}}({\vec x}_{1}, {\vec x}_{2})  = g^{2}\int\int K_{\Delta_{\phi}}(X, {\vec x}_{1})\,{\rm Tr}[{\widetilde S}_{\Delta_{\psi}}(X,Y)\,{\widetilde S}_{\Delta_{\chi}}(Y, X)]\,K_{\Delta_{\phi}}(Y, {\vec x}_{2}).
\end{equation}
Surely, to use this expression in spectral representation (\ref{44}) the replacements $\Delta_{\psi} \to d/2 + ic$, $\Delta_{\chi} \to d/2 + i{\overline c}$ must be performed in it.

Substitution in (\ref{45}) of split representations (\ref{20}) of ${\widetilde S_{\Delta_{\psi}}}$, ${\widetilde S_{\Delta_{\chi}}}$ gives two spinor-spinor-scalar vertices of type $\rm I$ (\ref{35}) with their convolution over two boundary points ${\vec x}_{a}$, ${\vec x}_{b}$. Thus the RHS of (\ref{45}) takes the form:

\begin{eqnarray}
\label{46}
g^{2} {\rm Tr} \int\int d{\vec x}_{a}\,d{\vec x}_{b} \, \left[\int \,K_{\Delta_{\phi}}(X, {\vec x}_{1})\,{\overline \Sigma}^{IR}_{\Delta_{\chi}}(X, {\vec x}_{b})\,\Sigma^{IR}_{\Delta_{\psi}}(X, {\vec x}_{a})\, dX \,\right] \, \cdot \qquad  \nonumber
\\   \nonumber
\\   \nonumber
\cdot \, \left[\int\,{\overline \Sigma}^{UV}_{d - \Delta_{\psi}}(Y,{\vec x}_{a})\,\Sigma^{UV}_{d - \Delta_{\chi}}(Y,{\vec x}_{b})\,K_{\Delta_{\phi}}(Y,{\vec x}_{2})\,dY\right] =  \qquad  \qquad \qquad \nonumber
\\
\\
= g^{2}\,{\rm Tr}\,\int\int d{\vec x}_{a}\,d{\vec x}_{b}\, {\cal M}^{{\rm 3 pt \, (I)}\,{\overline {IR}}-IR}_{\Delta_{\chi},\Delta_{\psi},\Delta_{\phi}}({\vec x}_{b},{\vec x}_{a},{\vec x}_{1})\,{\cal M}^{{\rm 3 pt \, (I)}\,{\overline {UV}}-UV}_{d - \Delta_{\psi},d - \Delta_{\chi},\Delta_{\phi}}({\vec x}_{a},{\vec x}_{b},{\vec x}_{2}).  \nonumber
\end{eqnarray}

Using (\ref{35}) for ${\cal M}^{{\rm 3 pt \, (I)}\,{\overline {IR}}-IR}$ and similar expression for ${\cal M}^{{\rm 3 pt \, (I)}\,{\overline {UV}}-UV}$ with evident changes of arguments, taking into account:

$$
{\rm Tr}[{\vec \gamma}({\vec x}_{a} - {\vec x}_{b})\,\Pi_{-}{\vec \gamma}({\vec x}_{a} - {\vec x}_{b})\,\Pi_{+}] = P_{ab}\,{\rm Tr}\Pi_{+} = \frac{P_{ab}}{2} \cdot dim\gamma
$$
($dim\gamma = {\rm Tr}[{\hat 1}]$ is equal to $d$ for $d$ even and equal to $(d - 1)$ for $d$ odd), and performing standard conformal integral (\ref{26}) over ${\vec x}_{b}$  it is obtained from (\ref{45}):

\begin{eqnarray}
\label{47}
{\cal H}^{{\rm 2pt}(0|\frac{1}{2}\,\frac{1}{2})}_{\Delta_{\phi}|\Delta_{\psi}\Delta_{\chi}}({\vec x}_{1}, {\vec x}_{2}) = \frac{g^{2} \, dim\gamma}{2}\, \frac{1}{P_{12}^{\Delta_{\phi} - \frac{d}{2}}}\, \int \frac{d{\vec x}_{a}}{P_{1a}^{\frac{d}{2}}P_{2a}^{\frac{d}{2}}}  \, \cdot   \qquad  \nonumber
\\   \nonumber
\\  \nonumber
\cdot \,   B^{\rm (I)}(\Delta_{\psi}, \Delta_{\chi}, \Delta_{\phi}; 1/2)\,B^{\rm (I)}(d - \Delta_{\psi}, d - \Delta_{\chi}, \Delta_{\phi}; 1/2) \, \cdot  \nonumber
\\
\\  \nonumber
\cdot\, A\left(\frac{\Delta_{\psi} - \Delta_{\chi} + \Delta_{\phi}}{2}, \frac{\Delta_{\chi} - \Delta_{\psi} + \Delta_{\phi}}{2}, d - \Delta_{\phi}\right),  \qquad
\end{eqnarray}
$A$, $B^{\rm (I)}$ see in (\ref{28}), (\ref{36}).

Typical for conformal theories divergent integral in (\ref{47}) was analyzed in \cite{Giombi1}; here the dimensional regularization is chosen when in general formulas (\ref{27}), (\ref{28}) it is taken

$$
d \to d^{*} = d + \epsilon, \, \, \, \beta_{1} = \beta_{2} = \frac{d}{2}.
$$ 

Leaving in integral in (\ref{47}) only most divergent term $\sim \epsilon^{-1}$ we absorb it in the "bare" coupling constant $g$ and define renormalized coupling as:

\begin{equation}
\label{48}
g_{R}^{2} = g^{2}\,P_{12}^{\frac{d}{2}}\, \int \frac{d{\vec x}_{a}}{P_{1a}^{\frac{d}{2}}P_{2a}^{\frac{d}{2}}} = g^{2}\, \frac{4\pi^{\frac{d}{2}}}{\Gamma(\frac{d}{2})} \, \frac{1}{\epsilon}.
\end{equation}

Thus using (\ref{48}) and deciphering $A$ (\ref{28}), $B^{\rm (I)}$ (\ref{36}) and ${\hat C}_{\Delta_{\psi, \, \chi}}$ (\ref{13}) that enter expression for $B^{\rm (I)}$ we get finally:

\begin{equation}
\label{49}
{\cal H}^{{\rm 2pt}(0|\frac{1}{2}\,\frac{1}{2})}_{\Delta_{\phi}|\Delta_{\psi}\Delta_{\chi}}({\vec x}_{1}, {\vec x}_{2}) = \frac{C_{\Delta_{\phi}}}{P_{12}^{\Delta_{\phi}}} \, \, g_{R}^{2} \, \, \frac{dim \gamma}{16 \pi^{d}} \, \, \frac{{\cal {\bf R}}(\Delta_{\psi}, \Delta_{\chi}; \Delta_{\phi})}{F^{(0)}(\Delta_{\phi})},
\end{equation}
where

\begin{equation}
\label{50}
F^{(0)}(\Delta_{\phi}) = \frac{\Gamma(\Delta_{\phi}) \, \Gamma(d - \Delta_{\phi})}{\Gamma\left(\Delta_{\phi} - \frac{d}{2}\right) \, \Gamma\left(\frac{d}{2} - \Delta_{\phi}\right)},
\end{equation}
and

\begin{eqnarray}
\label{51}
{\cal {\bf R}}(\Delta_{\psi}, \Delta_{\chi}; \Delta_{\phi}) = \frac{\Gamma\left(\frac{\Delta_{\psi} + \Delta_{\chi} + \Delta_{\phi} - d + 1}{2}\right) \, \Gamma\left(\frac{2d - \Delta_{\psi} - \Delta_{\chi} - \Delta_{\phi} + 1}{2}\right)}{\Gamma\left(\frac{1}{2} + \Delta_{\psi} - \frac{d}{2}\right) \, \Gamma\left(\frac{1}{2} + \frac{d}{2} - \Delta_{\psi}\right)} \, \cdot   \, \nonumber
\\   \nonumber
\\ \nonumber
\\ \nonumber
\cdot \, \frac{\Gamma\left(\frac{\Delta_{\psi} - \Delta_{\chi} + \Delta_{\phi}}{2}\right)\,\Gamma\left(\frac{\Delta_{\chi} - \Delta_{\psi} + \Delta_{\phi}}{2}\right) \, \Gamma\left(\frac{\Delta_{\psi} + \Delta_{\chi} - \Delta_{\phi} + 1}{2}\right)}{\Gamma\left(\frac{1}{2} + \Delta_{\chi} - \frac{d}{2}\right) \, \Gamma\left(\frac{1}{2} + \frac{d}{2} - \Delta_{\chi}\right)}  \, \cdot
\\  \nonumber
\\  \nonumber
\\
\cdot \, \frac{\Gamma\left(\frac{d + \Delta_{\psi} - \Delta_{\chi} - \Delta_{\phi}}{2}\right)\, \Gamma\left(\frac{d + \Delta_{\chi} - \Delta_{\psi} - \Delta_{\phi}}{2}\right) \, \Gamma\left(\frac{d - \Delta_{\psi} - \Delta_{\chi} + \Delta_{\phi} + 1}{2}\right)}{\Gamma\left(\frac{d}{2} - \Delta_{\phi}\right) \, \Gamma\left(1 + \Delta_{\phi} - \frac{d}{2}\right)},
\end{eqnarray}
or through bulk masses of spinors $m_{\psi} = \Delta_{\psi} - d/2$, $m_{\chi} = \Delta_{\chi} - d/2$ (see (\ref{12})) and Bessel functions' order $\nu = \Delta_{\phi} - d/2$ (see (\ref{5})) for scalar field:

\begin{eqnarray}
\label{52}
{\cal {\bf R}}(\Delta_{\psi}, \Delta_{\chi}; \Delta_{\phi}) = \frac{\Gamma\left(\frac{d}{4} + \alpha \right) \, \Gamma\left(\frac{d}{4} - \alpha \right) \, \Gamma\left(\frac{d}{4} + \beta \right) \, \Gamma\left(\frac{d}{4} - \beta \right)}{\Gamma\left(\frac{1}{2} + m_{\psi}\right) \, \Gamma\left(\frac{1}{2} - m_{\psi}\right)} \, \cdot  \nonumber
\\
\\
\cdot \, \frac{\Gamma\left(\frac{1}{2} + \frac{d}{4} + \gamma\right) \, \Gamma\left(\frac{1}{2} + \frac{d}{4} - \gamma\right) \, \Gamma\left(\frac{1}{2} + \frac{d}{4} + \delta\right) \, \Gamma\left(\frac{1}{2} + \frac{d}{4} - \delta\right)}{\Gamma\left(\frac{1}{2} + m_{\chi}\right) \, \Gamma\left(\frac{1}{2} - m_{\chi}\right) \, \Gamma(- \nu) \, \Gamma (1 + \nu)},  \nonumber
\end{eqnarray}

$$
\alpha = \frac{m_{\psi} - m_{\chi} + \nu}{2}, \, \, \qquad \beta = \frac{m_{\chi} - m_{\psi} + \nu}{2}, 
$$

$$
\gamma = \frac{m_{\psi} + m_{\chi} + \nu}{2}, \, \, \qquad \delta = \frac{m_{\psi} + m_{\chi} - \nu}{2}.
$$
${\cal {\bf R}}$ is the main function in our analysis. It can be seen that for even $d$ it is expressed in terms of elementary trigonometric or hyperbolic (for imaginary arguments, like in spectral representations (\ref{8}), (\ref{19})) functions. For example

for $d = 2$:
$$
{\cal {\bf R}}(\Delta_{\psi}, \Delta_{\chi}; \Delta_{\phi}) = - \frac{\pi \, \gamma \, \delta \, \cos\pi m_{\psi} \, \cos\pi m_{\chi} \, \sin\pi \nu}{\cos\pi \alpha \, \cos\pi \beta \, \sin\pi \gamma \, \sin\pi \delta}, \qquad \qquad \qquad \, \, \, \, \quad (52a)
$$

and for $d = 4$:
$$
{\cal {\bf R}}(\Delta_{\psi}, \Delta_{\chi}; \Delta_{\phi}) = - \frac{\pi  \alpha  \beta  \left(\frac{1}{4} - \gamma^{2}\right)  \left(\frac{1}{4} - \delta^{2}\right)  \cos\pi m_{\psi}  \cos\pi m_{\chi}  \sin\pi \nu}{\sin\pi \alpha \, \sin\pi \beta \, \cos\pi \gamma \, \cos\pi \delta} \, \, \quad (52b)
$$

As it could be expected, dependence on coordinates of harmonic bubble (\ref{49}), and hence of full bubble (\ref{44}), is the same like of the elementary scalar conformal correlator (\ref{7}), it is singled out in front of the RHS of (\ref{49}).

\vspace{0,5cm}

\subsection{Spinor bubble formed by spinor and scalar.}

\vspace{0,5cm}

Generated by the same bulk Yukawa coupling like in previous subsection one-loop 2-point contribution to the conformal correlator of spinor field $\psi (Z)$ when loop is formed by spinor field $\chi (Z)$ and scalar field $\phi (Z)$ has a form:

\begin{equation}
\label{53}
{\cal M}^{\rm {2pt\,(\frac{1}{2}|\frac{1}{2}\,0})}_{\Delta_{\psi}|\Delta_{\chi},\Delta_{\phi}}({\vec x}_{1},{\vec x}_{2}) = g^{2}\int\int\,{\overline \Sigma}^{IR}_{\Delta_{\psi}}(X, {\vec x}_{1})\,S_{\Delta_{\chi}}(X,Y)\,G_{\Delta_{\phi}}(X,Y)\,\Sigma^{IR}_{\Delta_{\psi}}(Y, {\vec x}_{2}),
\end{equation}
where spinor bulk-to-boundary propagators $\Sigma^{IR}_{\Delta_{\psi}}$ and ${\overline \Sigma}^{IR}_{\Delta_{\psi}}$ are given in (\ref{15}), (\ref{16}) and Green functions $G_{\Delta_{\phi}}$, $S_{\Delta_{\chi}}$ - in (\ref{8}), (\ref{19}) (IR option is meant, that is $\Delta_{\phi} > d/2$ and $\Delta_{\chi} > d/2$, see (\ref{5}), (\ref{12})).

Following the logic of previous section, with account of spectral representations of scalar (\ref{8}) and spinor (\ref{19}) Green functions and proportionality of corresponding harmonic functions to differences ${\widetilde G}$ (\ref{10}) and ${\widetilde S}$ (\ref{20}) of IR and UV scalar and spinor Green functions, double integral spectral representation of the one-loop spinor correlator (\ref{53}) is obtained:

\begin{equation}
\label{54}
{\cal M}^{{\rm 2pt}(\frac{1}{2}|\frac{1}{2}\,0)}_{\Delta_{\psi}|\Delta_{\chi}\Delta_{\phi}}({\vec x}_{1}, {\vec x}_{2}) = - \frac{1}{4\pi^{2}}\,\int\int \, \frac{dc\, {\overline c}\, d{\overline c} \, {\cal H}^{{\rm 2pt}(\frac{1}{2}|\frac{1}{2}\,0)}_{\Delta_{\psi}|\frac{d}{2} + ic, \frac{d}{2} + i {\overline c}}\,({\vec x}_{1}, {\vec x}_{2})}{[c + i\,(\Delta_{\chi} - \frac{d}{2})] \, [{\overline c}^{2} + (\Delta_{\phi} - \frac{d}{2})^{2}]},
\end{equation}
where harmonic bubble ${\cal H}^{{\rm 2pt}(\frac{1}{2}|\frac{1}{2}\,0)}$ in numerator is again built by the replacement in (\ref{53}) of two bulk Green functions $G$, $S$ with corresponding differences ${\widetilde G}$ (\ref{10}), ${\widetilde S}$ (\ref{20}):

\begin{equation}
\label{55}
{\cal H}^{{\rm 2pt}(\frac{1}{2}|\frac{1}{2}\,0)}_{\Delta_{\psi}|\Delta_{\chi}\Delta_{\phi}}({\vec x}_{1},{\vec x}_{2}) = g^{2}\int\int \, {\overline \Sigma}^{IR}_{\Delta_{\psi}}(X, {\vec x}_{1})\, {\widetilde S}_{\Delta_{\chi}}(X,Y)\,{\widetilde G}_{\Delta_{\phi}}(X, Y) \,\Sigma^{IR}_{\Delta_{\psi}}(Y, {\vec x}_{2})
\end{equation}
To use it in spectral representation (\ref{54}) the replacements $\Delta_{\chi} \to d/2 + ic$, $\Delta_{\phi} \to d/2 + i{\overline c}$ must be performed in (\ref{55}).

Substitution here of split representations of ${\widetilde G}$ (\ref{10}) and ${\widetilde S}$ (\ref{20}) gives for the RHS of (\ref{55}):

\begin{eqnarray}
\label{56}
g^{2} (d - 2\Delta_{\phi})\, \int\int d{\vec x}_{a}\,d{\vec x}_{b}\left[\int \,{\overline \Sigma}^{IR}_{\Delta_{\psi}}(X, {\vec x}_{1})\,\Sigma^{UV}_{d - \Delta_{\chi}}(X, {\vec x}_{a})\,K_{\Delta_{\phi}}(X, {\vec x}_{b})\,dX\right] \, \cdot   \qquad \nonumber
\\  \nonumber
\\ \nonumber
\cdot \, \left[\int\,{\overline \Sigma}^{IR}_{\Delta_{\chi}}(Y,{\vec x}_{a})\,\Sigma^{IR}_{\Delta_{\psi}}(Y,{\vec x}_{2})\,K_{d - \Delta_{\phi}}(Y,{\vec x}_{b})\,dY\right] = g^{2}(d - 2\Delta_{\phi})\, \cdot \qquad \qquad \qquad
\\ 
\\  \nonumber
\cdot \, \int d{\vec x}_{a} \int d{\vec x}_{b}\, {\cal M}^{{\rm 3pt \, (II)}\,{\overline {IR}}-UV}_{\Delta_{\psi},d - \Delta_{\chi},\Delta_{\phi}}({\vec x}_{1},{\vec x}_{a},{\vec x}_{b})\,{\cal M}^{{3pt \, \rm (I)}\,{\overline {IR}}-IR}_{\Delta_{\chi},\Delta_{\psi},d - \Delta_{\phi}}({\vec x}_{a},{\vec x}_{2},{\vec x}_{b}), \qquad \qquad \qquad 
\end{eqnarray}
here last equality follows from expressions (\ref{32}) and (\ref{37}) for ${\cal M}^{\rm 3pt \, (I)}$, ${\cal M}^{\rm 3pt \, (II)}$. After the evident changes of variables and arguments in final formulas (\ref{35})-(\ref{36}) and (\ref{40})-(\ref{42}) for these vertices, (\ref{56}) comes to:

\begin{eqnarray}
\label{57}
{\cal H}^{{\rm 2pt}(\frac{1}{2}|\frac{1}{2}\,0)}_{\Delta_{\psi}|\Delta_{\chi}\Delta_{\phi}}({\vec x}_{1}, {\vec x}_{2}) = B^{\rm (II)}(\Delta_{\psi},d - \Delta_{\chi},\Delta_{\phi}; 1/2) \, B^{\rm (I)}(\Delta_{\chi}, \Delta_{\psi}, d - \Delta_{\phi}; 1/2) \, \cdot \nonumber
\\
\\  \nonumber
\cdot \, g^{2} \, (d - 2\Delta_{\phi})\,\int d{\vec x}_{a}\, I_{b}({\vec x_{1}}, {\vec x_{2}}, {\vec x_{a}}) \, \frac{\Pi_{+}\,{\vec \gamma}({\vec x}_{2} - {\vec x}_{a})\,\Pi_{-}}{P_{1a}^{\delta^{\rm (II)}_{1a}}\,P_{2a}^{\delta^{\rm (I)}_{2a}}}, \qquad \qquad \qquad
\end{eqnarray}
where $B^{\rm (I)}$, $B^{\rm (II)}$ see in (\ref{36}), (\ref{41}) and $I_{b}({\vec x_{1}}, {\vec x_{2}}, {\vec x_{a}})$ is integral over ${\vec x_{b}}$:

\begin{eqnarray}
\label{58}
I_{b}({\vec x_{1}}, {\vec x_{2}}, {\vec x_{a}}) = \int d{\vec x}_{b} \, \frac{\left[- P_{1a} + P_{1b} + P_{ab} + ({\vec x_{1}} - {\vec x_{b}})^{\alpha} \, ({\vec x_{a}} - {\vec x_{b}})^{\beta} \, S^{\alpha\beta}\right]}{P_{1b}^{\delta^{\rm (II)}_{1b}}\,P_{2b}^{\delta^{\rm (I)}_{2b}}\,P_{ab}^{(\delta^{\rm (II)}_{ab} + \delta^{\rm (I)}_{ab})}} = \nonumber
\\ \nonumber
\\  \nonumber
= \frac{A\left(\delta^{\rm (II)}_{1b}, \delta^{\rm (I)}_{2b}, (\delta^{\rm (II)}_{ab} + \delta^{\rm (I)}_{ab})\right)\,\left(\frac{d}{2} - \delta^{\rm (II)}_{1b}\right)\,\left(\frac{d}{2} - (\delta^{\rm (II)}_{ab} + \delta^{\rm (I)}_{ab})\right)}{P_{12}^{\Delta_{\psi} + \frac{1}{2} - \frac{d}{2}} \, P_{1a}^{\frac{\Delta_{\phi} + \Delta{\chi} - \Delta_{\psi}}{2}}\, P_{2a}^{\frac{2d - \Delta_{\phi} - \Delta_{\psi} - \Delta_{\chi} + 1}{2}}} \, \cdot \qquad \quad \nonumber
\\  
\\
\cdot \, \left[- P_{1a} + P_{12} + P_{2a} - ({\vec x_{1}} - {\vec x_{a}})^{\alpha} \, ({\vec x_{2}} - {\vec x_{a}})^{\beta} \, S^{\alpha\beta}\right], \qquad \qquad \qquad \nonumber
\end{eqnarray}
$A$, $S^{\alpha \beta}$ are given in (\ref{28}), (\ref{31}). The values of six exponents $\delta_{ij}$ in (\ref{57})-(\ref{58}) corresponding to exponents in (\ref{36}) for vertex of type $\rm I$, and to exponents in (\ref{42}) for vertex of type $\rm II$ are here as follows:

\begin{eqnarray}
\label{59}
\delta_{2a}^{\rm (I)} = \frac{\Delta_{\psi}+ \Delta_{\chi} + \Delta_{\phi} + 1 - d}{2}; \, \, \, \, \, \, \delta_{2b}^{\rm (I)} = \frac{\Delta_{\psi} - \Delta_{\chi} - \Delta_{\phi} + d}{2}; \nonumber
\\
\\
 \delta_{ab}^{\rm (I)} = \frac{\Delta_{\chi} - \Delta_{\psi} + d - \Delta_{\phi}}{2}, \qquad  \qquad  \qquad \nonumber 
\end{eqnarray}
and

\begin{eqnarray}
\label{60}
\delta_{1a}^{\rm (II)} = \frac{\Delta_{\psi} + d - \Delta_{\chi} - \Delta_{\phi}}{2};\,\, \, \, \, \, \delta_{1b}^{\rm (II)} = \frac{\Delta_{\psi} +  \Delta_{\phi} - (d - \Delta_{\chi}) + 1}{2};\nonumber
\\
\delta_{ab}^{\rm (II)}  = \frac{(d - \Delta_{\chi}) + \Delta_{\phi} - \Delta_{\psi} + 1}{2}. \qquad  \qquad  \qquad
\end{eqnarray}

It is seen from (\ref{59}), (\ref{60}) that 

\begin{equation}
\label{61}
\delta^{\rm (II)}_{1b} + \delta^{\rm (I)}_{2b} + \delta^{\rm (II)}_{ab} + \delta^{\rm (I)}_{ab} = d + 1.
\end{equation}

In calculation of $I_{b}$ (\ref{58}) it was taken into account that conformal integrals over ${\vec x}_{b}$ corresponding to the second $(P_{1b})$ and the third $(P_{ab})$ terms in integrand of (\ref{58}) are the ordinary ones of type (\ref{26}), whereas sum of exponents of $P$ in the first $(P_{1a})$ term in the integrand in (\ref{58}) is, according to (\ref{61}), equal to $(d + 1)$; this integral is less trivial - see (\ref{29}). Integral over ${\vec x_{b}}$ of the last term in the RHS of (\ref{58}) also comes to standard conformal integral (\ref{26}) with account of asymmetry of $S^{\alpha\beta}$ and elementary identity

$$
\frac{({\vec x_{1}} - {\vec x_{b}})^{\alpha} \, ({\vec x_{a}} - {\vec x_{b}})^{\beta} \, S^{\alpha\beta}}{P_{ab}^{(\delta^{\rm (II)}_{ab} + \delta^{\rm (I)}_{ab})}} = - \frac{({\vec x}_{1} - {\vec x}_{a})^{\alpha} \, S^{\alpha\beta}}{2\,(\delta^{\rm (II)}_{ab} + \delta^{\rm (I)}_{ab} - 1)} \, \frac{\partial}{\partial x_{a}^{\beta}}\,\left[ \frac{1}{P_{ab}^{\delta^{\rm (II)}_{ab} + \delta^{\rm (I)}_{ab} - 1}}\right].
$$

Substitution of $I_{b}$ (\ref{58}) in (\ref{57}) with account of (\ref{59})-(\ref{61}) and (\ref{28}) gives:

\begin{eqnarray}
\label{62}
{\cal H}^{{\rm 2pt}(\frac{1}{2}|\frac{1}{2}\,0)}_{\Delta_{\psi}|\Delta_{\chi}\Delta_{\phi}}({\vec x}_{1}, {\vec x}_{2}) = \qquad  \qquad  \qquad  \qquad  \qquad \qquad \nonumber
\\ \nonumber
\\  \nonumber
= g^{2} \, (d - 2\Delta_{\phi})\, B^{\rm (I)}(\Delta_{\psi}, \Delta_{\chi},d - \Delta_{\phi}) \, B^{\rm (II)}(\Delta_{\psi}, d - \Delta_{\chi},\Delta_{\phi}) \, \frac{{\widetilde A}(\Delta_{\psi}, \Delta_{\chi}, \Delta_{\phi})}{P_{12}^{\Delta_{\psi} + \frac{1}{2} - \frac{d}{2}}} \, \cdot \nonumber
\\ 
\\ 
\cdot \, \int d{\vec x}_{a} \, \frac{[P_{2a} + P_{12} - P_{1a} - ({\vec x_{1}} - {\vec x_{a}})^{\alpha} \, ({\vec x_{2}} - {\vec x_{a}})^{\beta} \, S^{\alpha\beta}] \, {\vec\gamma}({\vec x}_{2} - {\vec x}_{a})\,\Pi_{-}}{P_{1a}^{\frac{d}{2}}\,P_{2a}^{\frac{d}{2} + 1}} \qquad \nonumber
\end{eqnarray}
where

\begin{equation}
\label{63}
{\widetilde A}(\Delta_{\psi}, \Delta_{\chi}, \Delta_{\phi}) = \frac{\pi^{\frac{d}{2}}\,\Gamma\left(\frac{2d - \Delta_{\psi} - \Delta_{\chi} - \Delta_{\phi} + 1}{2}\right) \, \Gamma\left(\frac{\Delta_{\chi} - \Delta_{\psi} + \Delta_{\phi}}{2}\right) \, \Gamma\left(\Delta_{\psi} + \frac{1}{2} - \frac{d}{2}\right)}{\Gamma\left(\frac{\Delta_{\psi} + \Delta_{\chi} + \Delta_{\phi} - d + 1}{2}\right) \, \Gamma\left(\frac{\Delta_{\psi} - \Delta_{\chi} + d - \Delta_{\phi}}{2}\right) \, \Gamma\left(d - \Delta_{\psi} + \frac{1}{2}\right)}.
\end{equation}

To take divergent conformal integrals over ${\vec x}_{a}$ in (\ref{62}) general formula (\ref{30}) is applied where according to dimensional regularization we change in the RHS $d \to d^{*} = d + \epsilon$ and put $\beta_{1}$, $\beta_{2}$ equal to corresponding values in every integral in (\ref{62}) (cf. in \cite{Giombi1}). This gives for the last line in (\ref{62}):

\begin{equation}
\label{64}
\int d{\vec x}_{a}\{...\} = \frac{{\vec \gamma}({\vec x}_{1} - {\vec x}_{2})\,\Pi_{-}}{P_{12}^{\frac{d}{2}}} \,\frac{2 \pi^{d/2}}{\Gamma(\frac{d}{2})} \, \frac{1}{\epsilon} \, \left( 1 + \frac{3}{d} \right)
\end{equation}

Final expression for spinor one-loop "harmonic bubble" (\ref{55}) is obtained from (\ref{62})-(\ref{64}) and expressions (\ref{36}), (\ref{41}) for coefficients $B^{\rm^(I)}$, $B^{\rm^(II)}$ with introduction of the renormalized coupling constant (\ref{47}) and with account that in general formulas (\ref{36}) and (\ref{41}) the corresponding replacements of $\delta_{ij}$ must be performed. Namely: $\delta_{12}^{\rm (I)}, \, \delta_{13}^{\rm (I)}, \, \delta_{23}^{\rm (I)}$ (in (\ref{36})) must be changed to $\delta_{2a}^{\rm (I)}, \, \delta_{ab}^{\rm (I)}, \, \delta_{2b}^{\rm (I)}$ given in (\ref{59}); and in the same way $\delta_{12}^{\rm (II)}, \, \delta_{13}^{\rm (II)}, \, \delta_{23}^{\rm (II)}$ (\ref{42}) must be changed to $\delta_{1a}^{\rm (II)}, \, \delta_{1b}^{\rm (II)}, \, \delta_{ab}^{\rm (II)}$ (\ref{60}). The result proves to be quite simple:

\begin{equation}
\label{65}
{\cal H}^{{\rm 2pt}(\frac{1}{2}|\frac{1}{2}\,0)}_{\Delta_{\psi}|\Delta_{\chi}\Delta_{\phi}}({\vec x}_{1}, {\vec x}_{2}) = {\hat C}_{\Delta_{\psi}} \, \, \frac{{\vec \gamma}({\vec x}_{1} - {\vec x}_{2}) \, \Pi_{-}}{P_{12}^{\Delta_{\psi} + \frac{1}{2}}} \, \, g_{R}^{2} \, \, \frac{\left(1 + \frac{3}{d}\right)}{32 \pi^{d}} \, \, \frac{{\cal {\bf R}}(\Delta_{\psi}, \Delta_{\chi}; \Delta_{\phi})}{F^{(\frac{1}{2})}(\Delta_{\psi})},
\end{equation}
where 

\begin{equation}
\label{66}
F^{(\frac{1}{2})}(\Delta_{\psi}) = \frac{\Gamma\left(\Delta_{\psi} + \frac{1}{2}\right) \, \Gamma\left(d - \Delta_{\psi} + \frac{1}{2}\right)}{\Gamma\left(\Delta_{\psi} + \frac{1}{2} - \frac{d}{2}\right) \, \Gamma\left(\frac{d}{2} - \Delta_{\psi} + \frac{1}{2}\right)},
\end{equation}
and coefficient ${\cal {\bf R}}(\Delta_{\psi}, \Delta_{\chi}; \Delta_{\phi})$ given in (\ref{51}) or (\ref{52}) is common for both harmonic bubbles ${\cal H}$ (\ref{49}), (\ref{65}) calculated in subsections 4.1, 4.2. Space-time and spinor dependence in front of the RHS of (\ref{65}) (and hence of full bubble (\ref{54})) copies the same of primary correlator (\ref{13}).

As it was noted above harmonic bubbles (\ref{49}), (\ref{65}) may be used as numerators of the integrands in spectral representations (\ref{44}) and (\ref{54}) after the corresponding replacements in universal function ${\cal {\bf R}}$ (\ref{51}) of conformal dimensions of intermediate fields by the integration variables $d/2 + ic$, $d/2 + i {\overline c}$. It is seen from expression for ${\cal {\bf R}}$ (\ref{51})-(\ref{52}) that double integrals (\ref{44}), (\ref{54}) are divergent in directions $c + {\overline c} = const$ and $c - {\overline c} = const$ \cite{Giombi1}.

In the next section the natural subtraction of divergences of bubbles (\ref{43}), (\ref{53}) will be performed.

\section{"Double-trace" elimination of UV-infinities}

\qquad The UV divergence of the self-energy diagrams reflected in particular in infinity of spectral integrals (\ref{44}), (\ref{54}) is a conventional difficulty in QFT. Here to overcome this difficulty we propose to apply to the bubble diagrams the double-trace from UV to IR flow approach used in \cite{Mitra} - \cite{Diaz2} for the unambiguous UV finite calculations of tadpoles and quantum vacuum energies of scalar and spinor bulk fields in spaces of arbitrary dimensions. For example in $AdS_{5}$ ($d = 4$) expressions for tadpoles defined in this way are as follows (differences ${\widetilde G}$ and ${\widetilde S}$ of "IR" and "UV" Green functions are given in (\ref{10}), (\ref{20}), and we restore here value $k_{AdS}$ of the curvature of AdS space):

for scalar tadpole \cite{Mitra} - \cite{Alt2}:

$$
{\widetilde G}^{(d = 4)}_{\Delta}(Z,Z) = \frac{k^{3}_{AdS}}{12\pi^{2}} \,(\Delta - 1)(\Delta - 2)(\Delta - 3),
$$

and for spinor tadpole \cite{Allais}, \cite{Diaz2}:

$$
{\widetilde S}^{(d = 4)}_{\Delta_{\psi} = 2 + m}(Z,Z) = \frac{k^{4}_{AdS}}{3\pi^{2}} \,\left(m^{2} - \frac{1}{4}\right) \, \left(\frac{9}{4} - m^{2} \right).
$$

Thus the difference of two similar Witten diagrams built of the UV or IR bulk Green functions proves to be finite and well defined for tadpoles, and we shall show that it is also finite and well defined for the bubbles. 

Most generally this modification of quantum diagrams means that instead of standard quantum generation functional, symbolically:

\begin{equation}
\label{67}
Z [j;G] = ({\rm{Det}}G)^{-1/2}\, e^{L_{int}\left(\frac{\delta}{\delta j}\right)} \, e^{\left(\frac{1}{2}jGj\right)}
\end{equation}
(here , $G$, $L_{int}$, $j$ are the free field Green function, the interaction Lagrangian and the field's source correspondingly; in frames of the AdS/CFT correspondence in generation functional of the $n$-point boundary conformal correlators $j({\vec x})$ is equipped with the corresponding bulk-to-boundary propagator), 

the ratio

\begin{equation}
\label{68}
{\widetilde Z} [j;G^{UV}, G^{IR}] = \frac{Z [j;G^{UV}]}{Z [j;G^{IR}]} = \frac{({\rm{Det}}G^{UV})^{-1/2}\, e^{L_{int}\left(i\frac{\delta}{\delta j}\right)} \, e^{\left(\frac{1}{2}jG^{UV}j\right)}}{({\rm{Det}}G^{IR})^{-1/2}\, e^{L_{int}\left(i\frac{\delta}{\delta j}\right)} \, e^{\left(\frac{1}{2}jG^{IR}j\right)}}
\end{equation}
of two quantum functionals determined by Green functions ($G^{UV}$ and $G^{IR}$) possessing two different asymptotics at the horizon must be considered as quantum generation functional for Witten diagrams. General analysis of the ratios of quantum functionals of one and the same bulk dynamics and different boundary conditions is presented in \cite{Barv2010}.

This approach means in particular that self-energy correlators (\ref{43}) and (\ref{53}) are redefined as a difference ${\cal {\widetilde M}}$ (also marked with tilde) of conventional bubble diagrams built of the products of two "UV" and two "IR" Green functions correspondingly:

\begin{eqnarray}
\label{69}
{\cal {\widetilde M}}^{{\rm 2pt}(0|\frac{1}{2}\,\frac{1}{2})}_{\Delta_{\phi}|\Delta_{\psi}\Delta_{\chi}}({\vec x}_{1}, {\vec x}_{2}) = {\cal M}^{{\rm 2pt}(0|\frac{1}{2}\,\frac{1}{2}) \, UV}_{\Delta_{\phi}|\Delta_{\psi}\Delta_{\chi}}({\vec x}_{1}, {\vec x}_{2}) - {\cal M}^{{\rm 2pt}(0|\frac{1}{2}\,\frac{1}{2}) \, IR}_{\Delta_{\phi}|\Delta_{\psi}\Delta_{\chi}}({\vec x}_{1}, {\vec x}_{2}) = \nonumber
\\
\\
= g^{2} \int\int K_{\Delta_{\phi}}(X ; {\vec x}_{1}) \, {\widetilde \Pi}^{{\rm 2pt}(0|\frac{1}{2}\,\frac{1}{2})}_{\Delta_{\psi}, \Delta_{\chi}}(X, Y) \, K_{\Delta_{\phi}}(Y ; {\vec x}_{2}) \, dX dY, \qquad  \qquad  \nonumber
\end{eqnarray}

\begin{eqnarray}
\label{70}
{\cal {\widetilde M}}^{\rm {2pt\,(\frac{1}{2}|\frac{1}{2}\,0})}_{\Delta_{\psi}|\Delta_{\chi},\Delta_{\phi}}({\vec x}_{1},{\vec x}_{2}) = {\cal M}^{\rm {2pt\,(\frac{1}{2}|\frac{1}{2}\,0}) \, UV}_{\Delta_{\psi}|\Delta_{\chi},\Delta_{\phi}}({\vec x}_{1},{\vec x}_{2}) - {\cal M}^{\rm {2pt\,(\frac{1}{2}|\frac{1}{2}\,0}) \, IR}_{\Delta_{\psi}|\Delta_{\chi},\Delta_{\phi}}({\vec x}_{1},{\vec x}_{2}) = \nonumber
\\
\\
= g^{2}\int\int\,{\overline \Sigma}^{IR}_{\Delta_{\psi}}(X, {\vec x}_{1}) \, {\widetilde \Pi}^{{\rm 2pt}(\frac{1}{2}|\frac{1}{2}\, 0)}_{\Delta_{\chi}, \Delta_{\phi}}(X, Y)  \,\Sigma^{IR}_{\Delta_{\psi}}(Y, {\vec x}_{2}) \, dX dY, \qquad \qquad  \nonumber
\end{eqnarray}
where

\begin{eqnarray}
\label{71}
{\widetilde \Pi}^{{\rm 2pt}(0|\frac{1}{2}\,\frac{1}{2})}_{\Delta_{\psi}, \Delta_{\chi}}(X, Y) = \{{\rm Tr}\,[S^{UV}_{\Delta_{\psi}} \, S^{UV}_{\Delta_{\chi}}] - {\rm Tr}\,[S^{IR}_{\Delta_{\psi}} \, S^{IR}_{\Delta_{\chi}}]\} (X, Y) = \qquad \nonumber
\\
\\
= \{ {\rm Tr} \, [{\widetilde S}_{\Delta_{\psi}} \, {\widetilde S}_{\Delta_{\chi}}] -  {\rm Tr}\,[S^{IR}_{\Delta_{\psi}} \, {\widetilde S}_{\Delta_{\chi}}] - {\rm Tr} \, [{\widetilde S}_{\Delta_{\psi}} \, S^{IR}_{\Delta_{\chi}}]\} (X, Y), \qquad \quad \nonumber
\end{eqnarray}

\begin{eqnarray}
\label{72}
{\widetilde \Pi}^{{\rm 2pt}(\frac{1}{2}|\frac{1}{2}\, 0)}_{\Delta_{\chi}, \Delta_{\phi}}(X, Y) = \{ S^{UV}_{\Delta_{\chi}}\,G^{UV}_{\Delta_{\phi}} - S^{IR}_{\Delta_{\chi}} \, G^{IR}_{\Delta_{\phi}}\}(X,Y) = \quad \qquad \nonumber
\\
\\
= \{{\widetilde S}_{\Delta_{\chi}} \, {\widetilde G}_{\Delta_{\phi}} - S^{IR}_{\Delta_{\chi}} \, {\widetilde G}_{\Delta_{\phi}} - {\widetilde S}_{\Delta_{\chi}} \, G^{IR}_{\Delta_{\phi}}\} (X, Y) \qquad \quad \nonumber
\end{eqnarray}
In derivation of (\ref{71}), (\ref{72}) identities $G^{UV} = G^{IR} - {\widetilde G}$ and $S^{UV} = S^{IR} - {\widetilde S}$ were used, where ${\widetilde G}$ and ${\widetilde S}$ are defined in (\ref{10}), (\ref{20})).

Then the divergent double integrals (\ref{44}), (\ref{54}) are canceled in (\ref{69}), (\ref{70}) and the remaining terms are UV-finite. With account of spectral and split representations of $G^{IR}$, ${\widetilde G}$ ((\ref{8}) - (\ref{10})) and of $S^{IR}$, ${\widetilde S}$ ((\ref{19})-(\ref{20})) it is seen that contribution of every of three terms of both ${\widetilde \Pi}$ (\ref{71}), (\ref{72}) to the expressions of the one-loop self-energy correlators (\ref{69}), (\ref{70}) includes corresponding harmonic bubbles (\ref{49}) and (\ref{65}). Thus finally for UV-finite bubbles (\ref{69}), (\ref{70}) defined according to prescription (\ref{68}) it is obtained:

\begin{eqnarray}
\label{73}
{\cal {\widetilde M}}^{{\rm 2pt}(0|\frac{1}{2}\,\frac{1}{2})}_{\Delta_{\phi}|\Delta_{\psi}\Delta_{\chi}}({\vec x}_{1}, {\vec x}_{2}) = \frac{C_{\Delta_{\phi}}}{P_{12}^{\Delta_{\phi}}} \, \, \frac{g_{R}^{2} \, dim\gamma }{16 \pi^{d} \, F^{(0)}(\Delta_{\phi})} \,\, \Biggl[{\cal {\bf R}}(\Delta_{\psi}, \Delta_{\chi}; \Delta_{\phi}) -   \qquad  \nonumber
\\
\\
- \int_{-\infty}^{+\infty} \frac{i \, dc}{2 \pi} \, \frac{{\cal {\bf R}}(\frac{d}{2} + i c, \Delta_{\chi}; \Delta_{\phi})}{[c + i (\Delta_{\psi} - \frac{d}{2})]} -\int_{-\infty}^{+\infty} \frac{i \, dc}{2 \pi} \, \frac{{\cal {\bf R}}(\Delta_{\psi}, \frac{d}{2} + i c; \Delta_{\phi})}{[c + i (\Delta_{\chi} - \frac{d}{2})]}\, \Biggr]. \qquad \nonumber
\end{eqnarray}
and

\begin{eqnarray}
\label{74}
{\cal {\widetilde M}}^{\rm {2pt\,(\frac{1}{2}|\frac{1}{2}\,0})}_{\Delta_{\psi}|\Delta_{\chi},\Delta_{\phi}}({\vec x}_{1},{\vec x}_{2}) = \qquad  \qquad  \qquad  \qquad \qquad  \qquad \nonumber
\\  \nonumber
\\  \nonumber
= \frac{{\hat C}_{\Delta_{\psi}} \, {\vec \gamma}({\vec x}_{1} - {\vec x}_{2}) \, \Pi_{-}}{P_{12}^{\Delta_{\psi} + \frac{1}{2}}} \, \frac{g_{R}^{2}\left(1 + \frac{3}{d}\right)}{32 \pi^{d} \, F^{(\frac{1}{2})}(\Delta_{\psi})} \,\, \Biggl[{\cal {\bf R}}(\Delta_{\psi}, \Delta_{\chi}; \Delta_{\phi}) -  \qquad  \qquad \qquad  \nonumber
\\
\\
- \int_{-\infty}^{+\infty} \frac{i \, dc}{2 \pi} \, \frac{{\cal {\bf R}}(\Delta_{\psi}, \frac{d}{2} + i c; \Delta_{\phi})}{[c + i (\Delta_{\chi} - \frac{d}{2})]} -\int_{-\infty}^{+\infty} \frac{i \, c \, dc}{2 \pi} \, \frac{{\cal {\bf R}}(\Delta_{\psi}, \Delta_{\chi}; \frac{d}{2} + i c)}{[c^{2} + (\Delta_{\phi} - \frac{d}{2})^{2}]}\, \Biggr], \qquad \qquad \nonumber
\end{eqnarray}
${\cal {\bf R}}$, $F^{(0)}$, $F^{(\frac{1}{2})}$ are given in (\ref{51})-(\ref{52}), (\ref{50}), (\ref{66}) correspondingly.

As noted above, function ${\cal {\bf R}}$ for even $d$ may be expressed in terms of elementary functions. In the next section bubbles (\ref{73}), (\ref{74}) will be written explicitly in the $d = 4$ Yukawa model.

\section{UV-finite bubbles in $SU(N)$ Yukawa model with conformal scalar field in $d = 4$}

\qquad Consider bulk Yukawa interaction of $N$ copies of spin $1/2$ fields $\psi (Z)$ of mass $m$ with the conformal invariant scalar field $\phi(Z)$ on $AdS_{5}$:

\begin{equation}
\label{75}
L_{int} = g \, \phi(Z)\,\Sigma_{k}{\overline \psi_{k}}(Z)\psi_{k}(Z).
\end{equation}

Conformal invariance of $\phi(Z)$ on $AdS_{d + 1}$ means that equation for scalar field includes curvature term $\xi_{c}\phi^{2}R_{d + 1}$ with coefficient $\xi_{c} = (d - 1)/4d$, and that in this case order parameter $\nu = \Delta_{\phi} - d/2 = 1/2$ in any dimension. Thus in this model $m_{\psi} = m_{\chi} = m$, $\nu = 1/2$ and it follows for the universal function ${\cal {\bf R}}$ (\ref{52}) in case $d = 4$:

\begin{equation}
\label{76}
{\cal {\bf R}}\left(2 + m, 2 + m; 2 + \frac{1}{2}\right) = \frac{\pi}{4} \, \frac{\cos^{2}\pi m}{\cos2\pi m} \, \left(m^{2} - \frac{1}{16}\right) \, \left(m^{2} - \frac{9}{16}\right) \equiv {\cal {\bf {\widetilde R}}}(m).
\end{equation}

Correspondingly for ${\cal {\bf R}}$ that enter the spectral integrals in the final expressions for bubbles (\ref{73}), (\ref{74}) it is obtained from (\ref{52}):

\begin{eqnarray}
\label{77}
{\cal {\bf R}}_{(d = 4)}\left(2 + ic, 2 + m; 2 + \frac{1}{2}\right) = {\cal {\bf R}}_{(d = 4)}\left(2 + m, 2 + ic; 2 + \frac{1}{2}\right) = \qquad \nonumber
\\
\\  \nonumber
=  - \frac{\pi}{8} \, \frac{\cos\pi m \, \cosh\pi c}{\cosh 2\pi c + \cos 2\pi m} \, \left[\left(m - \frac{1}{2}\right)^{2} + c^{2} \right] \, \left[\left(m + \frac{1}{2}\right)^{2} + c^{2} \right] \, \cdot \nonumber
\\   \nonumber
\\   \nonumber
\cdot \, \left(\frac{9}{4} - m^{2} + c^{2} - 2 i m c \right), \quad  \nonumber
\end{eqnarray}

\begin{eqnarray}
\label{78}
{\cal {\bf R}}_{(d = 4)}(2 + m, 2 + m; 2 + ic) = - \frac{i \pi \, c^{2} \, \sinh\pi c \cos^{2}\pi m}{(\cosh\pi c - 1)(\cosh\pi c + \cos 2\pi m)} \, \cdot \nonumber
\\
\\
\cdot \, \left[\left(m - \frac{1}{2}\right)^{2} + \frac{c^{2}}{4} \right] \, \left[\left(m + \frac{1}{2}\right)^{2} + \frac{c^{2}}{4} \right]. \nonumber 
\end{eqnarray}

Despite the complex nature of (\ref{77}), (\ref{78}) and of the integrands in spectral integrals in (\ref{73}), (\ref{74}), replacing there integration over $c$ from $- \infty$ to $+ \infty$ with integration from $0$ to $+ \infty$ will obviously give real RHS of (\ref{73}), (\ref{74}). 

Having in mind that in the Yukawa model under consideration ($d = 4$, $\Delta_{\psi} = \Delta_{\chi} = 2 + m$, $\Delta_{\phi} = 2 + 1/2$) we have $dim \gamma = 4$ and (see (\ref{50}) and (\ref{66})):

$$
F^{(0)}(\Delta_{\phi} = 5/2) = - \frac{3}{16}; \qquad F^{(\frac{1}{2})}(\Delta_{\psi} = m + 2) = \left(m^{2} - \frac{1}{4}\right) \, \left(m^{2} - \frac{9}{4}\right), 
$$
substitution of (\ref{76})-(\ref{78}) in (\ref{73}), (\ref{74}) gives final expressions for these one-loop conformal correlators (spinor-spinor and spinor-scalar bubbles):

\begin{eqnarray}
\label{79}
{\cal {\widetilde M}}^{{\rm 2pt}(0|\frac{1}{2}\,\frac{1}{2})\,(d = 4)}_{\Delta_{\phi} = 5/2|2 + m, 2 + m}({\vec x}_{1}, {\vec x}_{2}) = \frac{C_{5/2}}{P_{12}^{5/2}} \, \, g_{R}^{2} \, N \, \Phi^{(\phi)}(m),  \qquad  \nonumber
\\
\\
\Phi^{(\phi)}(m) =  \frac{4}{3 \pi^{4}} \Biggl[ - {\cal {\bf {\widetilde R}}}(m) - \frac{1}{4}\, m\, \cos\pi m \, I^{(1)}(m)\Biggr]  \qquad \nonumber
\end{eqnarray}
(there is multiplier $N$ in the RHS because in (\ref{75}) scalar field $\phi(Z)$ interacts with every of spinor fields $\psi_{k}(Z)$), 
and 

\begin{eqnarray}
\label{80}
{\cal {\widetilde M}}^{{\rm 2pt}(\frac{1}{2}|\frac{1}{2}\,0)\,(d = 4)}_{2 + m|2 + m, 5/2}({\vec x}_{1}, {\vec x}_{2}) = 
\frac{{\hat C}_{2 + m} \, {\vec \gamma}({\vec x}_{1} - {\vec x}_{2}) \, \Pi_{-}}{P_{12}^{2 + m + \frac{1}{2}}} \, g_{R}^{2} \, \Phi^{(\psi)}(m),  \qquad \nonumber
\\
\\
\Phi^{(\psi)}(m) = \frac{\left(1 + \frac{3}{4}\right)}{32 \pi^{4}} \,\, \frac{\Biggl[ {\cal {\bf {\widetilde R}}}(m) + \frac{1}{8}\, m\, \cos\pi m \, I^{(1)}(m)  - \cos^{2}\pi m \, I^{(2)}(m)\Biggr]}{\left(m^{2} - \frac{1}{4}\right) \, \left(m^{2} - \frac{9}{4}\right)}, \nonumber
\end{eqnarray}
where ${\cal {\bf {\widetilde R}}}(m)$ see in (\ref{76}), and

\begin{eqnarray}
\label{81}
I^{(1)}(m) =  \qquad \qquad \qquad  \qquad  \qquad \qquad \nonumber
\\
\\
= \int_{0}^{+\infty} \frac{dc \, \cosh\pi c \, \left(\frac{9}{4} - m^{2} + 3 c^{2}\right) \, \left[\left(m - \frac{1}{2}\right)^{2} + c^{2} \right] \, \left[\left(m + \frac{1}{2}\right)^{2} + c^{2} \right]}{(c^{2} + m^{2}) \, (\cosh 2\pi c + \cos 2\pi m)}, \nonumber
\end{eqnarray}

\begin{equation}
\label{82}
I^{(2)}(m) =  \int_{0}^{+\infty} \frac{dc \, c^{3} \, \sinh\pi c \, \left[\left(m - \frac{1}{2}\right)^{2} + \frac{c^{2}}{4} \right] \, \left[\left(m + \frac{1}{2}\right)^{2} + \frac{c^{2}}{4} \right]}{(c^{2} + \frac{1}{4}) \, (\cosh\pi c - 1) \, (\cosh\pi c + \cos 2\pi m)}
\end{equation}
are well defined convergent definite integrals.

Surely transparent expressions similar to (\ref{76}) - (\ref{82}), although somewhat more lengthy, may be put down for the model (\ref{75}) in case $d = 4$ when conformal dimension of scalar field $\Delta_{\phi} = 2 + \nu$ is arbitrary.

\section{"Old" conformal bootstrap in the AdS/CFT context: spectral equation for bulk spinor mass in Yukawa model of Sec. 6}

\qquad Expressions for UV finite quantum one-loop spinor-scalar contributions (bubbles) to scalar (\ref{73}) (or (\ref{79}) in Yukawa model (\ref{75})) and spinor (\ref{74}) ((\ref{80}) in model (\ref{75})) boundary-boundary conformal correlators may be used in different ways: 

- for calculation of anomalous dimensions generated by the scalar-spinor bulk loops like it was done in \cite{Giombi1} for bubbles built of the fields of integer spin with use of logarithmic terms in the dimensional regularization of conformal divergent integrals like (\ref{48}) or (\ref{64});

- for presenting decomposition of the bubbles built of two spinors or of spinor and scalar in an infinite series of residues in poles of integrands in spectral integrals in (\ref{73}), (\ref{74}); location and residues of these poles are evident from expressions (\ref{51})-(\ref{52}) (or (\ref{77})-(\ref{78}) in model (\ref{75})) for  coefficient ${\bf R}$ determining harmonic bubbles (\ref{49}), (\ref{65}); detailed analysis of poles of harmonic bubbles formed by the fields of integer spin was presented in \cite{Giombi1}, etc.

Here we pay attention that expressions obtained above permit to formulate spectral equations for the bulk masses of spinors in frames of the "old" conformal bootstrap in the AdS/CFT context \cite{Alt3}, \cite{Alt4}. The simplest "old" conformal bootstrap equation for Green function $G(X, Y)$ traditionally written in planar approximation looks as \cite{old1} - \cite{Grensing}: 

\begin{equation}
\label{83}
G (X_{1}, X_{2}) = g^{2} \, \int\int\, G(X_{1}, X)\,G(X, Y) \, G(X, Y) \, G(Y, X_{2})\, dX\,dY
\end{equation}
(triple interaction is supposed, $g$ is the coupling constant). Equating of exact Green functions to the one-loop quantum contribution built of the same exact Green functions is the main postulate of the "old" conformal bootstrap. This type of bootstrap equations was also used recently for calculation of spectra of conformal dimensions in one-dimensional Sachdev-Ye-Kitaev and in $d$-dimensional field theory models \cite{Klebanov}.

Eq. (\ref{83}) is just a conventional Schwinger-Dyson equation where terms associated with "bare" Lagrangian are omitted. This may be called the zero-Lagrangian approach applied by Sakharov in his quantum induced theory of gravity \cite{Sakharov} (the attempt to follow this way in the AdS context was made in \cite{Alt1}, \cite{Alt2} where UV-finite induced gravitational and gauge coupling constants were calculated). The similar “bootstrap” equations with zero “bare mass” terms are widely used in different approaches to dynamical symmetry breaking and mass generation pioneered in \cite{Jona}, \cite{Candelas}. 

In the AdS/CFT context it is assumed in \cite{Alt3}, \cite{Alt4} that $X_{1,2}$, $X$, $Y$ in (\ref{83}) are the bulk coordinates in $AdS_{d + 1}$ and $X_{1, 2}$ are sent to the horizon. This procedure \footnote{I am grateful to Ruslan Metsaev for this observation.} transforms LHS of (\ref{83}) into conformal correlator of the boundary conformal theory ((\ref{7}) for scalar and (\ref{13}) for spinor), whereas $G(X_{1}, X)$, $G(Y, X_{2})$ in the RHS become the corresponding bulk-to-boundary propagators. Thus RHS of (\ref{83}) becomes the quantum one-loop correlator (bubble). In this way different bootstrap equations may be obtained depending on the choice of boundary conditions for $G(X_{1}, X)$, $G(Y, X_{2})$ in (\ref{83}); in what follows we use "both IR" b.c. for these Green functions ("both UV" option will give the same spectral equations, as it was shown in \cite{Alt4} where model of interacting scalar fields was considered).

Applying "old" bootstrap equation (\ref{83}) (where $X_{1, 2}$ are sent to the horizon) to scalar and spinor fields and {\it defining} UV-finite bubble in the RHS of (\ref{83}) according to the "double trace subtraction" postulate of Sec. 5 we obtain from (\ref{83}) with account of (\ref{7}), (\ref{13}) - for the LHS, and (\ref{73}), (\ref{74}) - for the RHS:

\begin{eqnarray}
\label{84}
\frac{C_{\Delta_{\phi}}}{P_{12}^{\Delta_{\phi}}} = {\cal {\widetilde M}}^{{\rm 2pt}(0|\frac{1}{2}\,\frac{1}{2})}_{\Delta_{\phi}|\Delta_{\psi}\Delta_{\chi}}({\vec x}_{1}, {\vec x}_{2}); \qquad \nonumber
\\
\\
\frac{{\hat C}_{\Delta_{\psi}} \, {\vec \gamma}({\vec x}_{1} - {\vec x}_{2}) \, \Pi_{-}}{P_{12}^{\Delta_{\psi} + \frac{1}{2}}} = {\cal {\widetilde M}}^{\rm {2pt\,(\frac{1}{2}|\frac{1}{2}\,0})}_{\Delta_{\psi}|\Delta_{\chi},\Delta_{\phi}}({\vec x}_{1},{\vec x}_{2}). \nonumber
\end{eqnarray}

Reducing the similar space-time and spinor dependence of the LHS and RHS of Eq-s (\ref{84}) the following "old" bootstrap equations are obtained from (\ref{84}) for the particular case of model (\ref{75}) when bubbles are given by (\ref{79}), (\ref{80}):

\begin{equation}
\label {85}
1 = g_{R}^{2} \, N \, \Phi^{(\phi)}(m),
\end{equation}

\begin{equation}
\label {86}
1 = g_{R}^{2} \, \Phi^{(\psi)}(m).
\end{equation}

After elimination here of $g_{R}^{2}$ the interesting spectral equation for spinor bulk mass $m$ is obtained:

\begin{equation}
\label {87}
N \, \Phi^{(\phi)}(m) = \Phi^{(\psi)}(m).
\end{equation}

In \cite{Alt4} similar spectral equation for conformal dimensions in $O(N)$ symmetric model of $N$ scalar fields interacting with the conformal invariant Hubbard-Stratonovich field was derived, and its roots obeying unitarity bound demand were found: three roots for every $N = 1, 2, 3, 4$.

Spectral equation (\ref{87}) for every $N$ possesses positive and negative roots, however only one of them belongs to physically interesting interval (\ref{3}), and its value $m \cong 0.87$ weakly depends on $N$. According to (\ref{2}) this value of bulk fermion mass gives mass of physical fermion excitation $p_{0} \cong 0.05$ MeV.

The interesting task would be to consider spontaneous breakdown of $SU(N)$ symmetry in model (\ref{75}). This means that every spinor field $\psi_{k}(Z)$ is supposed to have its own bulk mass $m_{k}$ and in expression (\ref{79}) for the 2-spinor bubble of scalar field $\phi(Z)$ the multiplication by $N$ should be replaced by the sum over $k$: $N\, \Phi^{(\phi)}(m) \to \Sigma_{k}\Phi^{(\phi)}(m_{k})$, whereas expression (\ref{80}) should be valid for every $m_{k}$. Then spectral Eq. (\ref{87}) takes a form of system of $N$ equations ($k = 1, 2...N$) for $N$ unknowns $m_{k}$ ($\Phi^{(\phi, \psi)}(m)$ see in (\ref{79}), (\ref{80})):

\begin{equation}
\label {88}
\Sigma_{i}\Phi^{(\phi)}(m_{i}) = \Phi^{(\psi)}(m_{k})
\end{equation}

The physically intriguing goal is the search for roots $m_{1}... m_{N}$ of system (\ref{88}) that belong to interval (\ref{3}) with a hope to get, from Eq. (\ref{2}) for example, the observed masses of spin $1/2$ flavors.

Surely assumption of conformal invariance of scalar field in Yukawa model (\ref{75}) is just a demonstrative one. It is not difficult to put down the system of spectral equations similar to (\ref{88}) for arbitrary conformal dimension of scalar field $\Delta_{\phi}$ and to get the dependence of roots $m_{k}$ on $\Delta_{\phi}$. With a hope that for some $\Delta_{\phi}$ and $N$ the physically interesting combinations of $N$ roots $m_{1}... m_{N}$ will be obtained.

\section{Conclusion}

\qquad The results of this paper are three-fold, where in every direction there is possible further development.

First. The transparent expressions are obtained in physical $AdS_{d + 1}$ space for spinor-scalar vertices and for 2-point one-loop quantum conformal correlators (bubbles) formed with participation of spin 1/2 bulk fields. If the spinor-spinor bubble of scalar field was earlier calculated in the formalism of embedding space \cite{Carmi}, the result for the loop formed by spinor and scalar is not met in literature, as to our knowledge. The fantastic simplicity of this result obtained through lengthy calculations gives rise to the hunch that perhaps there is a simpler way to achieve it. The option of Yukawa bulk interaction $\phi(Z)\,{\overline \psi}(Z)\psi(Z)$ was considered above, and immediate generalization of approach of the paper may be the calculation of vertices and bubbles in the cases of pseudoscalar $\pi(Z)\,{\overline \psi}(Z) \gamma^{5} \psi(Z)$ or vector $V_{\mu}(Z)\,{\overline \psi}(Z) \gamma^{\mu} \psi(Z)$ interactions, the bulk QED as a special option.

Second. The novel tool of elimination of UV divergence of bubble Witten diagrams is proposed that actually repeats the approach used earlier in \cite{Mitra}-\cite{Diaz2} in calculations of Witten tadpoles. It would be interesting to check up if the difference of UV and IR Witten triangle and other diagrams proves to be UV-finite like it happened in Sec. 5 for bubble diagrams.

Third. The expressions for UV-finite spinor-scalar bubbles received in the paper permit to apply them in the "old" conformal bootstrap equations aimed at calculation of spectra of bulk spinor masses. The task for future may be the calculation of roots of spectral Eq. (\ref{87}) or of system of Eq-s (\ref{88}) and of similar equations received in models with pseudoscalar or vector interactions of spin 1/2 fields. As it was outlined in the Introduction the knowledge of bulk spinor masses may be a way to solution of the longstanding FMH (Fermion Mass Hierarchy, or Flavors Mass Hierarchy) problem.

\section*{Acknowledgements} Author is grateful to Ruslan Metsaev, Mitsuhiro Nishida and Kotaro Tamaoka, Dmitry Nesterov, Alexander Smirnov, Arkady Tseytlin and Mikhail Vasiliev for stimulating comments and assistance, and to participants of the seminar in the Theoretical Physics Department of the P.N. Lebedev Physical Institute for useful questions.

\section{Appendix. Derivation of formulas (\ref{24}) and (\ref{29})}

\subsection{Derivation of $D^{(1)}$ (\ref{24})}

\qquad Calculation of bulk integral $D^{(\alpha)}_{\gamma_{1}\gamma_{2}\gamma_{3}}({\vec x}_{1}, {\vec x}_{2}, {\vec x}_{3})$ (\ref{21}) is performed in an ordinary way with introduction of Schwinger parameters - see for example Section 3.1 and Appendix A in \cite{Paulos}. The representation of $D^{(\alpha)}$ as integral over Schwinger parameters differes from the well known one for $D^{(0)}$ only in the argument of the first Gamma-function and in the additional factor $(\Sigma_{i}t_{i})^{\alpha}$ in the integrand ($i = 1, 2, 3)$:

\begin{equation}
\label {89}
D^{(\alpha)}_{\gamma_{1}\gamma_{2}\gamma_{3}}({\vec x}_{1}, {\vec x}_{2}, {\vec x}_{3}) = \pi^{\frac{d}{2}} \, \frac{\Gamma\left(\frac{\Sigma_{i}\gamma_{i} - d - \alpha^{-1}}{2}\right)}{\Gamma(\gamma_{1}) \, \Gamma(\gamma_{2}) \, \Gamma(\gamma_{3})} \, \int\, \Pi_{i}\left(\frac{dt_{i}}{t_{i}} \, t_{i}^{\gamma_{i}}\right)\, (\Sigma_{i}t_{i})^{\alpha}\, e^{-Q^{2}},
\end{equation}
where 

$$
Q^{2} = t_{1}t_{2}P_{12} + t_{1}t_{3}P_{13} + t_{2}t_{3}P_{23},
$$
for $P_{ij}$ see (\ref{6}). 

Thus in case $\alpha = 1$ the standard change in (\ref{89}) of the integration variables

$$
t_{1} = \sqrt\frac{m_{2}m_{3}}{m_{1}} \qquad t_{2} = \sqrt\frac{m_{1}m_{3}}{m_{2}} \qquad t_{3} = \sqrt\frac{m_{1}m_{2}}{m_{3}}
$$
and simple algebra immediately give for $D^{(1)}$ expression (\ref{24}).

\subsection{Derivation of $R^{(1)}$ (\ref{29})}

\qquad Derivation of $R^{(1)}_{\beta_{1}\beta_{2}\beta_{3}}({\vec x}_{1}, {\vec x}_{2}, {\vec x}_{3})$ is again performed in an ordinary way with introduction of Schwinger parameters, that gives:

\begin{equation}
\label {90}
R^{(1)}_{\beta_{1}\beta_{2}\beta_{3}}({\vec x}_{1}, {\vec x}_{2}, {\vec x}_{3}) = 2\, \pi^{\frac{d}{2}} \, \int\, \Pi_{i}\left(\frac{dt_{i}}{t_{i}}\frac{t_{i}^{\beta_{i}}}{\Gamma(\beta_{i})}\right)\, (\Sigma_{i}t_{i})^{\Sigma_{i}\beta_{i} - d}\, e^{-Q^{2}},
\end{equation}
$Q^{2}$ see above in Sec. 9.1. For $\Sigma_{i}\beta_{i} = d + 1$ integral in (\ref{90}) coincides with the integral in (\ref{89}) at $\alpha = 1$ and (\ref{29}) follows from (\ref{90}) in the same way like (\ref{24}) was obtained from (\ref{89}).

\end{document}